\documentclass[12pt]{article}
\usepackage[utf8]{inputenc}

\usepackage[top=1in, right=1in, bottom=1in, left=1in]{geometry}
\usepackage{verbatim}
\usepackage{appendix}
\usepackage{graphicx}
\usepackage{cite}
\usepackage{float}
\usepackage[round]{natbib}
\usepackage{amsmath, amssymb, amsthm}
\usepackage{latexsym, marvosym}
\usepackage[usenames,dvipsnames]{xcolor}
\usepackage{enumerate}
\usepackage{enumitem}
\usepackage{array}
\usepackage{booktabs}
\usepackage[bottom]{footmisc}
\usepackage{pdfpages}
\usepackage{wrapfig}
\usepackage[colorlinks=true, citecolor=blue]{hyperref}
\usepackage{adjustbox}
\usepackage{multirow}
\usepackage{multicol}
\usepackage{mleftright}
\usepackage{tikz}
\usepackage[T1]{fontenc}
\usepackage{lmodern}
\usepackage{courier}
\usepackage{pifont}
\usepackage{microtype}
\usepackage{pdfpages}
\usepackage{setspace}
\usepackage{lipsum}
\usepackage{lscape}
\usepackage{subcaption}
\usepackage{hyperref}
\usepackage{footmisc}
\hypersetup{
    colorlinks=true,
    linkcolor=blue
}

\begin{document}
\begin{doublespacing}

%%%%%% Title Page
%%% Title Page
\begin{titlepage}
\title{Individual Heterogeneity and Cultural Attitudes in Credence Goods Provision}
\author{Johnny Jiahao Tang\footnote{I am grateful for helpful comments from Edward Glaeser, Andrei Shleifer, Ben Enke, Nathan Nunn, Alberto Alesina, Neil Thakral, Nihar Shah, Greg Bruich, Yueran Ma, Scott Kominers, Natalia Rigol, Hongyao Ma, and seminar participants. All errors are my own. jtang01@g.harvard.edu}\\Harvard University}

\date{First Version: March 28, 2018\\Current Version: \today}
\maketitle
\begin{abstract}

I study the heterogeneity of credence goods provision in taxi drivers taking detours in New York City. First, I document that there is significant detouring on average by drivers. Second, there is significant heterogeneity in cheating across individuals, yet each individual's propensity to take detours is stable: drivers who detour almost always detour, while those who do not detour almost never do. Drivers who take longer detours on each trip also take such trips more often. Third, cultural attitudes plausibly explain some of this heterogeneity in behavior across individuals.

%I use a methodology of detecting cheating by taxi drivers who take anomalous detours to earn higher fares to study the behaviors and determinants of cheating. I document that while there is significant detouring on average by drivers, there is significant heterogeneity in such behavior across individuals. Each individual's propensity to cheat is stable: drivers who detour almost always detour, while those who do not detour almost never do. Cultural norms plausibly explain some of this heterogeneity in cheating across individuals, while tracking individuals over time suggests that such heterogeneity in cheating is inconsistent with learning-by-doing.

% RETURN TO OLD ABSTRACT STYLE

\bigskip
\end{abstract}
\setcounter{page}{0}
\thispagestyle{empty}
\end{titlepage}
\pagebreak \newpage

%%%%%% Introduction
\section{Introduction}
Information asymmetries exist in many markets where sellers are able to identify the quality and quantity of service needed but the customers cannot. These markets are referred to as credence goods markets (\cite{dulleck2006}), and they are found in many economic settings: healthcare provisions (\cite{gruber1996}, \cite{gruber1999}), \cite{iizuka2007}), auto repairs (\cite{wolinsky1993}, \cite{wolinsky1995}), and taxi rides (\cite{balafoutas2013}, \cite{balafoutas_kerschbamer_sutter_2017}) are three common examples of such settings. Taxi drivers, for example, may provide a longer trip than is optimal because they may be financially incentivized to do so, whereas their passengers may not know the local roads and traffic conditions to identify the optimal route. Such behaviors by sellers in these settings are commonly referred to as cheating.

In spite of the prevalence of these transactions, little is known about how such cheating behaviors vary across individuals in these real-world economic interactions. In particular, is cheating widespread in the population, or is willingness to engage in such behavior higher for some individuals than others? Furthermore, if people have different propensities to cheat, where does this heterogeneity come from? Answering these questions requires a setting with the following four components: (1) a situation where people may cheat, (2) a credible methodology of detecting cheating, (3) repeated observations of the same individuals cheating, and (4) background information on these individuals. In this paper, I study a setting where all four components are present: taxi drivers in New York City.

% I study cheating and its determinants at the individual level in a large and broad population: taxi drivers in New York City.

%***Previous papers??? This is because it is challenging to collect data on and detect individuals cheating in the real world. Answering these questions requires a setting with the following four components: (1) a situation where people may cheat, (2) a credible methodology of detecting cheating, (3) repeated observations of the same individuals cheating, and (4) background information on these individuals. In this paper, I identify a setting where all four components are present: taxi drivers in New York City.

% \footnote{Taxi drivers taking passengers on detours is a common example of cheating covered in popular press around the world (see Times of India, Istanbul Insider, and Daily Mail \hyperref[references]{articles} referenced). It is also in line with the standard definition of cheating: “to act dishonestly or unfairly in order to gain an advantage.”}

% Specifically, I compare taxi drivers’ detouring behaviors on trips from LaGuardia airport to midtown Manhattan, a route on which they are paid by distance driven and time and thus incentivized to take detours, to the same driver’s behavior on trips from John F. Kennedy airport to the same destination in midtown Manhattan, a route on which they are paid a flat rate with no incentive to detour. 

This paper identifies cheating by taxi drivers who take anomalous detours that result in higher fares. I compare taxi drivers’ detouring behaviors on trips from LaGuardia airport to Manhattan, where they are paid by distance driven and thus incentivized to take detours, to the same drivers' behavior on trips from John F. Kennedy airport to the same destination in Manhattan, where they are paid a flat rate with no incentive to detour. Applying this methodology to over 430,000 taxi trips by 24,506 taxi drivers in 2013, I ask the following questions: first, how much cheating on average is there amongst taxi drivers in this setting? Second, if there is cheating on average, to what extent is there heterogeneity across individuals in their propensities to cheat? And lastly, if such differences across individuals exist, where does it come from?

% My paper documents three main findings on these questions. First, there is significant detouring on average by drivers. These detours do not save time or avoid traffic on average, and the same drivers do not detour when there is no financial incentive to do so. Second, there is significant heterogeneity in cheating at the individual level, yet each individual's propensity to cheat is stable: drivers who detour almost always detour, while those who do not detour almost never do. Drivers who take longer detours on each trip also take such cheating trips more often. Third, cultural norms plausibly explain some of this heterogeneity in cheating across individuals.

I first document cheating on average by taxi drivers on LaGuardia to Manhattan trips. Taxi rides from LaGuardia airport to Manhattan take an average detour of 5.6\% of the trip distance. Furthermore, the average level of detouring is driven by a subset of trips that have significant detours, with 32.4\% of trips detouring 10\% of the distance or more. One plausible explanation for detours is that drivers take detours to save passengers time or due to unobserved traffic conditions. I test this explanation by comparing the time durations of different taxi trips from LaGuardia airport going to the same block in Manhattan at the same time. I find that trips that took detours also took longer time and thus did not save passengers time. To further document this behavior as consistent with cheating, I find that the average detour distance is significantly lower on flat rate JFK airport to Manhattan trips. The average detour distance on JFK to Manhattan trips is only 1.5\% of the trip distance and only 8.3\% of trips have detours of 10\% of the distance or more.

Motivated by this finding, I analyze the extent to which individuals differ in their propensities to cheat through taking detours. I calculate the average amount of detouring each driver takes on all his or her LaGuardia-to-Manhattan trips in 2013. I find that there is significant heterogeneity in the amount of cheating across individuals. The standard deviation of drivers' average detour amount is 7.5\% of the average trip distance, compared to an average detour distance of 5.6\%. I then examine whether propensities to cheat for any given individual is stable across time by comparing the detour behavior of a driver on all trips he made in the first six months of 2013 to the same driver's detour behavior on all trips he made in the last six months of 2013. \autoref{persistence_insample_LGALGA_time_LGA1HLGA2H} shows that detour behavior is highly persistent across time: the correlation between the same driver's average detour distances across the two halves of 2013 is 0.76 (t-stat = 96.86). Placebo tests using drivers' flat rate JFK-to-Manhattan trips show that there is no correlation between a driver's detour behavior on LaGuardia trips and the same driver's detour behavior on JFK trips. \autoref{persistence_insample_LGAJFK_time_LGA1HJFK2H} shows that the correlation between the same driver's average detour distance on LaGuardia-to-Manhattan trips and on JFK-to-Manhattan trips is -0.02 (t-stat = -0.49). Combined, the findings suggest that there is significant heterogeneity in cheating across individuals, yet each individual's cheating type is stable. Reconciling this heterogeneity with the average level of cheating, the observed average level of cheating appears to be driven by a subset of individuals who consistently and predictably engage in such behavior, rather than cheating being a behavior observed across the entire population.

Individual heterogeneity in the amount of cheating on each trip also induces a behavioral response: drivers who engage in more cheating on each LaGuardia-to-Manhattan trip also make these trips more often. An increase of 100 more LaGuardia-to-Manhattan trips per year is associated with an increase of 1.06 miles in the average detour distance per trip, equivalent to 10.7\% of the average trip distance. A corresponding placebo test using JFK-to-Manhattan trips show that the number of JFK trips a driver makes is not associated with the amount of detouring on each JFK-to-Manhattan trip. This behavioral response also suggests that heterogeneity in the severity of cheating leads to differential frequencies of cheating in a setting where individuals can choose both how often they engage in cheating behavior and how much they cheat each time.

In the last section, I analyze why cheating behavior differs across individuals despite facing the same incentive structures in identical market settings. There are two main hypotheses. One hypothesis is self-selection, in which individuals have intrinsically different propensities to cheat due to different attitudes about cheating. Another hypothesis is learning, in which individuals learn to cheat through experience, such as a case where drivers may learn to cheat after making the same trip many times, so that differences in cheating is a reflection of differences in experience levels of drivers.

I test for the attitudes-based self-selection hypothesis by considering whether an individual's behavior is associated with the cultural attitudes of their home country. The idea is that cultural attitudes observed and internalized by individuals in their home countries may continue to influence their behavior even when they are no longer in settings where those attitudes are prevalent (\cite{fisman_miguel_2007}, \cite{barr2010}). Taxi drivers in New York City, 94\% of whom were born outside the US and hail from over 96 different countries\footnote{Based on New York City Taxi and Limousin Commission administrative data.}, provide an ideal population to test this hypothesis. The drivers come from countries which range in corruption norms and as such should exhibit differential behaviors under the attitudes-based hypothesis.

I exploit the quasi-random assignment of taxi drivers to trips at each airport to estimate the effect of drivers' home country cultural attitudes on the detour distances of each trip. At each airport, taxi drivers are sequentially assigned to passengers in line based on the orders in which their taxicabs and passengers arrived in their respective queues. Once in this queuing assignment, drivers cannot choose passengers based on observable characteristics or destination, both by law and in practice. As such, if cultural attitudes do indeed affect drivers' decisions to take detours, then the amount of detouring on each trip being correlated with the cultural attitudes of the driver's predicted home countries would be consistent with the cultural attitudes hypothesis. Furthermore, given the findings on the behavioral response of drivers, under the cultural attitudes hypothesis, drivers' home countries' norms should also be correlated with their frequencies of making detouring trips.

I find evidence consistent with this attitudes-based hypothesis. Drivers who are likely from countries with higher corruption levels cheat more and make LaGuardia trips more often. I match individual drivers' names to a predicted set of home countries using NamePrism, an academic name classification algorithm\footnote{NamePrism is trained on 74 million names from 118 countries that classifies names to predicted countries of origin.}. I then use Transparency International's Corruption Perceptions Index (CPI) as a measure of cultural attitudes in the home countries of the drivers and calculate a weighted average corruption norms score for each individual driver's predicted set of likely home countries. I find that an increase in the CPI score of 50 points (for example, from Nicaragua (28) to Germany (78)) is associated with a decrease in the average detour distance of the driver by 0.21 miles, compared to an average detour distance of 0.42 miles across all LaGuardia-to-Manhattan trips. The same increase in CPI score is also associated with 1.5 fewer LaGuardia to Manhattan trips and 2.1 more JFK to Manhattan trips per driver per year. Placebo tests using flat rate JFK trips show that higher corruption levels in predicted home countries are not associated with longer detours or making JFK trips more frequently.

On the other hand, I find evidence inconsistent with the learning hypothesis. I exploit the panel structure of the trip data to track how individuals' detouring behavior change over time. If heterogeneity in cheating is due to individuals having different experience levels, then drivers should exhibit differential detouring behaviors as they learn as they make more trips. I find that drivers do not change their detour behaviors with experience over time. Drivers who take detours in earlier trips continue to take detours in subsequent trips, while drivers who do not take detours in earlier trips do not take detours in subsequent trips. I repeat this analysis in both the full set of drivers as well as restricting to just new drivers, for whom the learning effects, if they exist, may plausibly be the strongest, and find no evidence of such effects in both samples.

This paper contributes to the literatures on cheating and credence goods and cultural attitudes along several dimensions.
Prior work on cheating have studied behaviors both in field settings and more recently in laboratory experiments. Previous field studies have found that people tend to cheat in aggregate or lacked repeated observations of individuals to study heterogeneities in cheating across individuals (\cite{fisman_miguel_2007},  \cite{jacob_levitt_2003}, \cite{martinelli_parker_perez-gea_rodrigo_2018}, \cite{heron_lie_2007}, and \cite{efendi_srivastava_swanson_2007}). In credence goods markets, cheating has been documented in car mechanics (\cite{wolinsky1993}, \cite{wolinsky1995}), healthcare providers (\cite{gruber1996}, \cite{gruber1999}), and taxi drivers (\cite{balafoutas2013}, \cite{balafoutas_kerschbamer_sutter_2017}). Recent laboratory experiments including \cite{dulleck2011} and \cite{kerschbamer2017social} have studied cheating in provision of goods in credence goods games. In particular, \cite{kerschbamer2017social} have been able to directly measure individual heterogeneities in preferences in a credence goods game in the laboratory setting, which is a challenge to do in field settings like the one in this paper. They derive theoretical predictions of how subjects would behave as sellers under different social preferences and identify the subjects’ social preferences in the credence goods game through their choices in the experiment. They find that subjects demonstrate significant heterogeneity in their social preferences.

Relative to the laboratory experiments and prior field findings, this paper contributes to the literature on cheating along several dimensions. First, this paper uses the panel structure of observational data from a real-world credence good market to document heterogeneity across individuals in cheating behavior in the field. In particular, this paper is complementary to the laboratory experiments of \cite{kerschbamer2017social} by studying a field setting in which individual heterogeneities may manifest in different behaviors by sellers. Second, this paper provides evidence that such individual heterogeneity may be linked to cultural attitudes. This is in part possible because of the rich variations in cultural attitudes of the home countries that NYC taxi drivers hail from. Third, this paper uses the time-series of the observations to test theories of learning in cheating behavior and find evidence inconsistent with learning. Fourth, the institutional setting of this paper offers several advantages for studying individual behavior in cheating: the incentives in this setting are material and significant components of the subjects’ income. Driving is a main source of income for over 80\% of taxi drivers\footnote{Figure 19 in Schaller (2004). Main source of income defined as working over 20 hours a week. 67\% of all drivers are full-time drivers, defined as working over 40 hours a week.} and cheating profits are a material percentage of their incomes\footnote{See \hyperref[magnitude_of_cheating]{Section 4.3.2} for a discussion on the magnitudes of cheating.}. Additionally, the field setting is clean and well-identified. All New York City taxi drivers drive identical yellow cabs on the same streets of New York City and face the same wage and fare structures. Furthermore, taxi drivers offer an attractive population to study as they are real expert sellers in a credence good market that is less likely to suffer from external validity concerns. Lastly, the setting of this paper is large scale, including over 24,000 drivers and over 430,000 taxi rides, which allows for precise tests to distinguish between different explanations of cheating.

The setting of this paper also fits in works on taxi drivers cheating that are well-documented in the literature. \cite{balafoutas2013} and \cite{balafoutas_kerschbamer_sutter_2017} use field experiments in Greece to study how passenger characteristics affect driver cheating. They find that taxi drivers both over-charge passengers by applying incorrect tariffs and over-treat passengers by taking them on detours. \cite{balafoutas2013} documents an average detour length of 10\% (1.3 km out of 12.7 km average ride) in the Athens taxi rides, which is slightly higher than the 5.6\% average detour in my New York City setting. Concurrent work by \citeauthor{liu2018} (forthcoming) study how passenger characteristics affect driver cheating using the NYC airport setting to show drivers cheat non-local passengers, while \cite{liu_brynjolfsson_dowlatabadi_2018} find that ride-sharing monitoring technologies for passengers may mitigate driver cheating. This paper contributes to this literature by studying driver characteristics rather than passenger characteristics in determining cheating. In particular, driver characteristics are related to the two main novel findings in this paper relative to prior work on taxi drivers cheating, that there is substantial heterogeneity in cheating behaviors across individual drivers, and that cultural attitudes may provide a plausible mechanism for this heterogeneity, whereas the evidence is inconsistent with explanations of learning-by-doing.

Finally, this paper connects to the literature on the persistent effects of cultural attitudes on economic behavior. \cite{fisman_miguel_2007} find that corruption norms are associated with parking violations by foreign diplomats in the US. \cite{barr2010}, \cite{gächter_schulz_2016} provide experimental evidence that cultural norms are related to willingness to engage in cheating and corruption in experimental games. This paper contributes to the literature by studying a real-world setting that is large scale, where subjects face identical incentives and decision problems, where the incentives are a significant source of subjects' incomes, and where the subjects are an arguably more generalized population in taxi drivers rather than top foreign diplomats or elite university students.

The rest of the paper proceeds as follows. Section 2 describes the study design and data. Section 3 documents baseline cheating across all drivers. Section 4 identifies individual heterogeneities in cheating and provides evidence on the behavioral response of drivers. Section 5 discusses cultural attitudes and learning as plausible causes of heterogeneity and tests predictions of each hypothesis. Section 6 concludes.

%%%%%% Study Design and Data
\section{Study Design and Data}
This paper uses data on taxicab trips made by licensed taxi drivers in New York City. This dataset has been previously studied in the labor supply literature on whether drivers set income targets which dictate their supply of labor (\cite{camerer1997}, \cite{farber2015you}, \cite{thakral2017}). This setting and dataset present several advantages to studying individual heterogeneity in cheating. First, taxi trips are a setting where cheating may occur (see e.g. \cite{balafoutas2013}). Second, the rich data detailing precise trip times, pickup and dropoff locations, and other characteristics allow for identification of cheating over other alternative explanations for detours. Third, driver-level data on each trip allows for repeated observations of the same individuals’ cheating behaviors. Fourth, driver registration data can link the individuals’ driving behaviors to background information about the drivers.

The data sample I use is all trips that originated from LaGuardia airport or JFK airport and ended in midtown Manhattan. Importantly, taxi drivers queue at each airport and are quasi-randomly assigned passengers in the order in which they and the passengers queue. Drivers also cannot by law refuse passengers based on observable characteristics, including the passenger's destination within Manhattan. I compare how drivers alter their route choices on LaGuardia-to-midtown trips, which are paid by distance, to the same drivers' route choices on JFK-to-midtown trips, which are fixed fare. \autoref{fig:map} provides a map of the airports and Manhattan areas.

\subsection{Taxi Fare Structure}
The yellow taxicabs are regulated taxis which operate under a pre-determined fare schedule set by New York City’s Taxi and Limousine Commission (TLC). The TLC is the government agency that regulates all taxicabs and for-hire vehicles in New York City. The TLC standard rate for a New York City taxi ride is a base fare of \$2.50 plus \$0.50 per every 1/5 mile driven or 60 seconds spent driving in traffic (below 12 mph) plus any additional surcharges\footnote{Given the fact that detour distances are largely due to distances driven on the highways into Manhattan, the distance-based detour cost is the primary determinant of the higher fare.}. Surcharges are usually fixed and include items such as a 30-cent improvement surcharge. I restrict my sample to the year 2013 since the year's TLC data contains the most complete information and it is the first full year after a rate change in 2012\footnote{The rate prior to the increase was \$0.40 per mile and \$45 for the JFK flat fare.}. Over 95\% of trip fares paid by credit card also included a non-zero tip, with the three most common percentages across all NYC taxi rides being 20\% (44\% of rides), 25\% (10\% of rides), and 30\% (4\% of rides)\footnote{Tips paid in cash are rarely recorded in TLC trip data. See \cite{haggag2014default} for a discussion on tipping in NYC taxis.}.

Trips from LaGuardia airport to midtown Manhattan follow this standard fare structure. Trips from JFK Airport, on the other hand, are a \$52 flat fare plus surcharges. In both cases, passengers are also responsible for paying for all tolls from crossing bridges and tunnels. Taxi drivers require a medallion license to operate the vehicle. Drivers who own a medallion do not pay any additional costs to drive, while drivers who do not own a medallion usually pay a fixed cost to rent the medallion for each shift they drive\footnote{Each vehicle is usually in operation for two shifts per day. The morning shift usually begins around 6:30-7:30AM and the night shift usually begins at 5-6PM. See: NYC TLC Taxi Factbook.}. All fares and tips earned from the trips are kept by the driver.

\subsection{Driver Detouring Considerations}
On any given trip, a driver makes a decision on whether to take a detour and if so, to what extent, from the pickup location to the destination. In this section, I discuss some considerations, including the benefits and costs, for the driver in making decisions to take detours. 

The main benefit to taking a detour is higher earnings. Consider a driver who takes an average detour of 2 miles per trip. On each trip, the driver will earn an additional \$6\footnote{Based on \$2.50 per mile and a 20\% default tip (see e.g. \cite{haggag2014default}).}, which is an increase of 19.6\% on the average fare amount of a LaGuardia-to-Manhattan taxi ride. A driver who takes 300 such LaGuardia to Manhattan in 2013 would make up to \$1,800 in additional earnings a year. This amounts to an extra 5.3\% in income for the median New York City taxi driver making \$34,020 per year\footnote{Source: US Bureau of Labor Statistics.}.

%Formally, passengers have the right to request a route of their choice, although in practice, exercising this choice depends on knowledge of local route conditions and awareness of the right itself, which passengers picked up at LaGuardia and JFK airports, who are likely to be non-locals, may not be knowledgeable of. In practice, drivers almost always make the decision on the route, unless passengers request a specific route, in which case if the requested route is reasonable, the driver is to follow the route.

There are several potential types of costs to the driver for deciding to take a detour. One is the costs associated with getting caught taking detours. Drivers could be questioned by the passenger during the trip or have passengers either refuse to tip or file complaints to the Taxi Limousine Commission that result in financial penalties after the fact. While it is difficult to directly observe the interactions within the taxi, complaints leading to legal enforcement on taxi driver cheating are very rare. Less than 0.001\% of all taxi rides have been subject to penalties and disciplinary actions due to drivers not following reasonable routes\footnote{Based on Taxi and Limousine Commission's administrative data. The maximum fine is \$150.}. Additionally, previous works on tipping including \cite{haggag2014default} find that tip amounts are strongly influenced by non-trip related factors including the default tip amounts presented. Consistent with this, I find that passengers do not appear to alter their tip amounts in response to drivers' detour behaviors.

A second type of costs is intrinsic costs associated with norms and psychological guilt. Drivers may have different attitudes and beliefs about the acceptability and prevalence of taking detours. In particular, drivers may either have different internal beliefs on whether taking detours is right or wrong, which could lead to internal psychological costs, or they may have different views about how prevalent or socially accepted it is. I explore the potential effects of attitudes-based types of costs on drivers' decisions to take detours in Section 5 of this paper.

%three main types of decisions during any given shift. The first decision is the locations to search for passengers. The second decision is deciding which route to take, once a passenger has been picked up and a destination has been given. The third decision is, after driving for a certain amount of time, when to end the shift. This study focuses on how cheating behavior is related to the first two types of decisions here. On deciding where to look for passengers, the driver can choose to search and pickup passengers anywhere within the five boroughs of New York City\footnote{This applies to yellow taxicabs only. There are restrictions on green taxicabs, known as "boro taxis", on pickups within Manhattan. For the purpose of this study, I only consider yellow taxicabs rides.}. Since 

\subsection{Data Source}
The main dataset is the New York City Taxi and Limousine Commission (TLC) tripsheet data. The dataset records information on every New York City yellow taxicab ride in 2013 through the taxicab’s payment meter and GPS systems. Each observation in the dataset is a single yellow taxicab ride that occurred in 2013. \autoref{summary-stats} presents summary statistics on the trip data. I also use Google Maps data to determine the recommended route based on the precise pick-up and drop-off locations and the exact time of travel. The recommended route is the route that Google Maps directs users to take when they request route guidance directions with the same pick-up and drop-off locations and time of travel as the taxi trip. I combine the TLC tripsheet data with TLC driver license registration data, which contains information on the name of the driver, license number, and license registration and expiry dates. 

In the cultural attitudes analysis, I use survey scores and indices from Transparency International’s corruption perceptions index (CPI) in 1998 and 2013, as a proxy for cultural attitudes in the home countries of each driver. \autoref{demographics-summary-stats} reports the number of all New York City taxi drivers by predicted home countries by name-origin group. I then match individual drivers to predicted home countries by name using the NamePrism classification software. For each driver, I compute the average CPI score of the predicted home countries, weighted by the total number of drivers from each country based on the TLC Factbook data, as a proxy of cultural attitudes of the driver’s home country and match it to the driver’s trip data. \autoref{demographics-countries-scores} reports the name-origin country groups and the CPI scores of each group. \hyperref[appendix_culture_data]{Appendix A} contains details on the construction of the dataset.

There are a few limitations to using Transparency International's CPI to measure cultural attitudes. First, cheating norms are not directly measured by perceptions of corruption. Several prior works, however, have sought to establish a link between cheating behavior and corruption. \cite{gächter_schulz_2016} use a measure of corruption to construct a general measure of prevalence of rule violations, which they link to cheating behaviors in experimental games across countries. \cite{barr2010} use Transparency International's CPI as a measure of corruption norms in an experimental bribery game to link corruption with dishonest behavior in the form of participants accepting bribes. \cite{hanna2017} link cheating in experimental games to propensity for corruption by government workers. These experimental evidences all relate corruption to some form of cheating or dishonest behavior, but inferring cheating attitudes from CPI may still remain an imperfect way of measuring cheating attitudes.

Second, cheating attitudes may also be correlated with other factors that may influence the propensities to take detours. Two potential confounding factors include differential demands for money and driver-passenger interactions. In Section 5.3.4, I attempt to test for the effects of differential demands for money by exploiting plausible time variations in demand for money across demographic groups to study how cheating behaviors differ during times of high money demand. I do not find support for such differential demand effects on cheating, although it is plausible that the effects may be present across demographic groups. On driver-passenger interactions, passengers may place different levels of scrutiny on drivers based on the observable characteristics of the drivers that may in turn be correlated with cheating attitudes in their home countries. The rarity of passenger complaints and potential information asymmetries facing non-local airport passengers, however, may suggest that the effects of passenger scrutiny are likely to be  diminished.

\subsection{Detour Distance and Time Measurements}
\subsubsection{Detour Distance}
I use three measures of trip distance to measure driver detour behavior. The first is the Trip Distance in miles, which is the trip distance as recorded by the taxicab’s meter system. The second is the Detour Distance, which for each trip is calculated as Detour Distance = (Recorded Trip Distance) / (Google Maps distance) – 100\%, where the Google Maps distance is the Google Maps distance of a counterfactual trip with the same origin, destination, time, and day of the week. This provides a first-differenced measure of trip distances as an independent control for time of day and destination variations. The third measure is in-sample detour distance, which is the additional distance driven on this trip compared to the average distance driven of all other similar trips in the sample. Similar trips are defined as all trips going to the same destination area that happen within a one-hour time window as this trip, excluding this trip. The grids are equally sized and constructed based on the reported GPS coordinates of the dropoff locations. The dropoff area I study is divided into 16 blocks and each corresponds approximately to a width of one avenue and a length of four streets (approximately 0.2 miles by 0.2 miles). For the in-sample detour distance, I remove observations for which there are fewer than 5 similar trips. Comparing trips to both Google Maps and other actual taxi trips at similar times from and to the same locations accounts for unobserved traffic conditions and allow for direct comparisons of trip distances and times.

One potential limitation to using the Google Maps benchmark is that it may not be able to account for idiosyncratic traffic conditions on a daily level. I address this by repeating the analysis in this paper with the in-sample detour distance, which has the ability to compare trips within the same day and hour, to identify heterogeneity in cheating behavior across individuals. I also compare drivers' routes from LaGuardia airport to Manhattan to their routes from a residential area just outside of LaGuardia to Manhattan to identify detouring in the aggregate\footnote{An ideal benchmark would be to use trips from the same hour of the same day from the residential area as a counterfactual to the LaGuardia trips. One data limitation is that residential area trips do not occur at a high enough frequency to construct such a benchmark (there are almost never more than 5 similar trips by such a criteria).}.

\subsubsection{Time}
I also use three analogous measures of trip time. The first is the trip time in minutes, as recorded by the taxicab’s meter system. The second is the Google Maps Time Discrepancy in minutes, which for each trip is calculated as Google Maps Time Discrepancy = (Recorded Trip Time) - (Google Maps trip time), where the Google Maps trip time is the Google Maps route time of a counterfactual trip with the same origin, destination, time, and day of the week. This provides a first-differenced measure of trip time as an independent control for time of day and destination variations. The third is in-sample time discrepancy, which is the additional time taken on this trip compared to the average time taken of all other similar trips in the sample, constructed analogous to in-sample detour distance. For the in-sample time discrepancy, I remove observations for which there are fewer than 5 similar trips. 

%%%%%% Baseline Cheating
\section{Baseline Cheating}
The first baseline result is that there is a significant amount of detouring for trips from LaGuardia airport and that these detours do not save passengers time. \autoref{detour-tables} compares the detour distances of taxi drivers on trips from LaGuardia airport and from JFK airport. The average detour distance of LaGuardia airport trips is 5.6\% of the trip distance (0.42 miles). Furthermore, there is a large number of trips with significant detours: 32\% of trips have detour distances of 10\% or more, and 8\% of trips have detour distances of 30\% or more. The high level of detouring on average and substantial number of trips with significant detours are both consistent with cheating by taxi drivers on these trips.

To further identify this as cheating, I compare the average detour behaviors of drivers on LaGuardia airport to Manhattan trips, which are paid by distance, to JFK airport to Manhattan trips, which are fixed rate. Consistent with a cheating explanation for taking detours, the average detour distance of fixed-rate JFK trips is only 1.5\% (0.23 miles). Despite LaGuardia to Manhattan trips being on average a little over half as long as JFK to Manhattan trips (9.9 miles from LaGuardia compared to 17.6 miles from JFK), the average miles detoured on LaGuardia to Manhattan trips is more than twice that of JFK trips (0.42 miles average detour from LaGuardia compared to 0.19 miles from JFK). Furthermore, the number of trips with significant detours is substantially lower at JFK: only 8\% of JFK trips have detour distances of 10\% or more, and almost no JFK trips have detour distances of 30\% or more. \autoref{detour_histogram_LGA_pct} and \autoref{detour_histogram_JFK_pct} present the distribution of trips in histogram format. The differential outcomes in driver routing behaviors are consistent with a cheating explanation of taking detours.

There are alternative explanations for why detouring may be due to non-cheating reasons, including detouring to save time, risk aversion, and drivers having wrong beliefs about optimal routes. I use the richness of the data to test these alternative explanations of cheating.

\subsection{Detours to save time?}
One plausible explanation for drivers taking detours is to save passengers time or due to unobserved traffic conditions. I test for whether detours save time or avoid traffic by comparing the time discrepancies between trips that took detours and trips that did not take detours. The regression I run is as follows:

$$Y_{i,jk} = \alpha_{1} + \alpha_{2} \cdot X_{i,jk} + \alpha_{3,jk} \cdot Z_{i,jk} + \epsilon_{i,jk}$$

Where $Y_{i,jk}$ is a measure of trip time in minutes of the i-th trip that goes to destination Grid k at time j. $X_{i,jk}$ is the detour distance in miles of the trip. $Z_{i,jk}$ is a set of controls for time of the trip, day of the trip, and destination of the trip. Trip time is denoted by time of day and day of week j (j = 1,2,...,168) and trip destination is denoted by grid number k (k = 1,2,...,16). The coefficient $\alpha_{2}$ tests for the relationship between trip time and detour distance. Standard errors are clustered at the driver level. \autoref{lga-time-regressions} reports the results of the regression of trip time on detour distance. For every additional mile of detour, trips take on average 0.38 minutes longer in recorded trip time, 0.44 minutes longer than the Google Maps predicted time, and 0.38 minutes longer than the average trip time of other trips going to the same destination at similar times. The coefficients are all statistically distinguishable from zero at the 1\% level. If drivers took detours to save time or avoid traffic, trips with longer detours than similar trips at the same time going to the same place should take shorter on average, not longer. As such, the results are inconsistent with the explanation that taxi drivers take detours to save passengers time or avoid traffic.

\subsection{Alternative explanations}
\subsubsection{Risk aversion}
Another alternative explanation is that drivers may be helping passengers minimize the uncertainty of their trip time if the detour takes longer on average but has less variance in time. For instance, consider two routes: one is a non-detour route that is on average 30 minutes but has a chance of being 60 minutes or longer. Another route is a detour route that is on average 40 minutes but is never longer than 50 minutes. Then it is plausible that passengers may actually prefer the detour route if they want to insure against a chance of a longer-than-60 minutes trip. One way to test for this explanation is to compare the variances of the trip times by detour distance. If trips with longer detour distances have lower variances of trip time, then it is possible that drivers take detours to insure passengers against the risk of a very long trip. If trips with longer detour distances have higher average times and greater variances of trip time, then it is inconsistent with this use of detours as an insurance against long trip times.

I compute the standard deviations of trip time discrepancies for trips by detour distance. The standard deviation of trip time discrepancies is 8.03 for trips with detours longer than 1-mile, 7.34 for trips with detours between -1 and 1 mile, and 7.81 for trips with detours less than -1 miles. The results are robust to using different cutoffs. As such, trips with longer detour distances do not have less variation in trip time, which is inconsistent with the explanation that drivers take detours because trips with detours have less variation in trip times.

\subsubsection{Left tail of detour distances}
Another plausible concern is that there are rides which took routes shorter than what is recommended by Google Maps. Shorter trips do not earn drivers more money, so it is unlikely that they are taking such trips for financial reasons. One explanation for this behavior is that these are trips where passengers request to take the toll-free, shortest route (via Queensboro bridge)\footnote{Passengers also have the right to request a route of their choice, although exercising this choice likely depends on knowledge of local route conditions and knowledge of the right itself, which non-locals may not be aware of.}. 4.83\% of LaGuardia trips in the right tail of the histogram (with a detour distance of greater than 1-mile) were toll-free, whereas 35.3\% of left tail trips (with a detour distance shorter than -1 mile) were toll-free, compared to the full sample of LaGuardia trips, of which 10.4\% of trips were toll-free. This suggests that the left tail trips were more likely to be toll-free trips and could plausibly have been requested by the passenger.

\subsubsection{Drivers having wrong beliefs about optimal routes}
One alternative explanation is that drivers have wrong beliefs about the best routes and are naive or benevolently mistaken in trying to take detours to save passengers time, but that they are systemically wrong in estimating trip times and end up taking detours that take longer in time. This explanation is at odds with several findings. First, at JFK airport, where there is no financial incentive to take detours, taxi drivers do not take detours. Instead, drivers take the shortest and fastest trips. Second, \autoref{yankee-stadium} shows detour distances of two types of trips: (1) trips from a residential area just outside of LaGuardia airport to midtown Manhattan, and (2) trips from Yankee Stadium to midtown Manhattan. For both sets of trips, the passengers are more likely to be locals and thus more knowledgeable about the routes taken. Consistent with a cheating explanation for detours, these trips take significantly shorter detours than trips from LaGuardia airport. In particular, the finding that trips from the residential area just outside of LaGuardia airport do not take detours is also inconsistent with traffic explanations for detours. Taken together, the finding that drivers do not take local passengers on detours, which is consistent with prior work on information asymmetry in the taxi market by \cite{balafoutas2013}, \cite{balafoutas_kerschbamer_sutter_2017}, and \citeauthor{liu2018} (forthcoming), is also inconsistent with wrong belief-based explanations for detours.

%%%%%% Individual Heterogeneity
\section{Individual heterogeneity in cheating}
Having established that there is cheating by taxi drivers on average, I now analyze differences in cheating behaviors across individuals. There are two main hypotheses. One hypothesis is that under identical incentives and institutional settings, individuals have the same propensities to cheat. Another hypothesis is that individuals have intrinsically different propensities to cheat, even in the same situations. Using repeated observations of the same drivers’ detouring behaviors over the course of many trips, I find evidence consistent with the hypothesis that there is significant heterogeneity in cheating at the individual level, yet each individual's propensity to cheat is stable: drivers who detour almost always detour, while those who do not detour almost never do.

I test for heterogeneity in cheating at the individual level using two empirical strategies. I first analyze the extent to which individuals differ in their propensities to cheat through taking detours. I calculate the average amount of detouring each driver takes on all his or her LaGuardia-to-Manhattan trips in 2013. I include only drivers who made at least 10 such trips in 2013 to avoid idiosyncratic variations driving the results. I present the results using in-sample detour distances to capture variation across drivers in the sample and report in the appendix that the results are quantitatively and qualitatively similar using the Google Maps detour distance. The results are also consistent if different minimum number of trips criteria are used. If drivers have homogeneous propensities to take detours, then the average detour distances should be relatively similar across drivers. This is because the detour distance by construction is the portion of trip distance that cannot be explained by the time, date, drop-off, and pickup locations, and is residualized to have mean 0 across all trips. As such, I take variations in average detour distances across drivers as evidence in favor of heterogeneous propensities to cheat and against homogeneous propensities.

In the second empirical strategy, I examine whether propensities to cheat for any given individual is stable across time by testing whether the detour distances of trips the driver made in an earlier observation sample predict the same driver’s detour distances on trips in a later observation sample. I use a driver’s trips in the first half of 2013 (January 1 through June 30, inclusive) as the early sample and the same driver’s trips in the second half of 2013 (July 1 to December 31, inclusive) as the later sample. If the propensity to take detours is constant across drivers, then the detour distance of the same driver should not be predictable across time, for the same reason as before that detour distance by construction is the portion of trip distance that cannot be explained by the time, date, drop-off, and pickup locations, and is residualized to have mean 0 across all trips. For the detour distance of a driver's trips to be correlated across rides over time, therefore, the driver’s propensity to take detours on each ride must also be correlated across rides.

I first find that there is significant heterogeneity in the amount of cheating across individuals. The standard deviation of drivers' average detour amount is 7.5\% of the average trip distance. This is compared to an average detour distance of 5.6\% across all trips. Previous work including \cite{balafoutas2013} on taxi driver cheating have been unable to detect the heterogeneity in drivers' propensities to engage in such cheating due to the lack of driver level information. The large standard deviation of average detour distances across drivers suggests that the there is substantial heterogeneity across drivers that is not captured in the average amount of cheating observed.

\autoref{persistence_insample_LGALGA_time_LGA1HLGA2H} presents the analysis in scatterplot form. Each data point represents an individual driver. The dispersion of average detour distances amongst drivers in each half of 2013 confirms that the overall level of heterogeneity across individual drivers persists across time. Furthermore, despite the heterogeneity in the amount of cheating across drivers, there is very little variation in cheating for any given driver across time. This is seen in the distinct and highly persistent positive correlation between a driver’s detour distances on trips in the first half of 2013 and the second half of 2013 0.76 (t-stat = 96.86). Together, the results strongly suggest that there is substantial heterogeneity in the propensity to take detours across different drivers and that this propensity is persistent for a given individual across time. This difference in propensities to take detours across individuals is consistent with individual heterogeneities in cheating and inconsistent with individuals having homogeneous propensities to cheat.

As a falsification test, I repeat this analysis comparing the detour behaviors of a driver’s LaGuardia to Manhattan trips to the same driver’s detour behaviors on JFK to Manhattan trips trips. If the persistence in detour behavior is indeed an indication of individual propensities to cheat, a driver’s detour behavior on LaGuardia-to-Manhattan trips should not be correlated with the same driver’s detour behavior on flat rate JFK-to-Manhattan trips, for there is no incentive to cheat on the latter type of trips.

\autoref{persistence_insample_LGAJFK_time_LGA1HJFK2H} presents the results of this falsification test. Consistent with detouring as an indication of propensities to cheat, there is no economically or statistically significant correlation in detouring behavior between LaGuardia- and JFK-to-midtown trips (correlation = 0.02, t-stat = 0.69). As such, the findings of the falsification test are consistent with individuals having heterogeneous propensities of cheating.

One potential concern with using in-sample detour distances as a benchmark is that if two trips took the exact same route but their benchmark trips were different, then the detour distances on the two trips would be different. To address this concern, I report in \autoref{persistence_gmap_LGALGA_time_LGA1HLGA2H} the results using Google Maps detour distances, which are quantitatively and qualitatively similar. Additionally, any such variation in the benchmark trips are likely to be informative of actual routing conditions at the time of the trip, such that the optimal route may change over time.

\subsection{Behavioral response}
One immediate consequence of this heterogeneity is that differences in individuals' average detour distances directly implies differences in the profitability of LaGuardia taxi trips. I document that increased profitability on LaGuardia trips induces a behavioral response in drivers by incentivizing them to go to LaGuardia more often. I show that drivers who have longer detour distances on average also make more LaGuardia trips. As a placebo test, I find that drivers’ detour distances on JFK trips are not related with the frequency of JFK trips. I interpret the findings as evidence of a behavioral response induced by the increased profitability of cheating behavior for drivers who take longer detours to engage in cheating behavior more often.

For each driver, I compute the total number of LaGuardia trips made by the driver in 2013. I then compare the number of LaGuardia trips made by a given driver to his average detour distances on these trips. \autoref{LGA_distance_numtrips} presents the relationship in scatterplot form, binned by the driver’s total number of LaGuardia trips in 2013. Columns 1 and 2 in \autoref{frequency_regs} present the regression results. A clear positive relationship exists between the average severity of a driver’s detours and the number of times a driver makes LaGuardia trips. The relationship is both statistically and economically significant. For every 100 additional LaGuardia trips a driver makes per year, his average detour distance is 1.01 miles longer (t-stat = 180.35), which is approximately 10\% of the total trip distance in the detour distance.

As a placebo test, I repeat the analysis by comparing the driver’s average detour distance on JFK trips to the number of JFK trips the driver makes in 2013. Since JFK trips are flat fare, the profitability of the trips do not vary with detour distances, and as such there should not be a behavioral response by the drivers at JFK. \autoref{JFK_distance_numtrips} and columns 3 and 4 in \autoref{frequency_regs} present the results of this analysis at JFK. Consistent with the behavioral response hypothesis, drivers do not exhibit differential frequencies of making JFK trips in response to differences in detour distances. The coefficient estimate on the relationship between number of JFK trips and distance of each trip is neither economically nor statistically significant.

One assumption this analysis makes is that drivers are able to autonomously choose which trips (e.g. airport or non-airport) they take and which airport they do go to. Both by TLC rules and in practice, drivers can freely choose where they work their trips. Taxi drivers' search decisions, particularly between airport and non-airport trips) are well-documented and studied in the literature (see e.g. \cite{buchholz2019})\footnote{Even within airport trips, taxi drivers in practice can choose which airports they want to serve by either taking trips near the airport and driving directly to the airport, or queuing at popular locations such as hotels or taxi stands where they can coordinate with the hotel doorman or taxi stand operator on where they want to take trips to.}. 

A related alternative explanation is that drivers may feel they are 'forced' to go to LaGuardia as opposed to JFK when searching for an airport trip, for example when they pickup a passenger at a hotel, and therefore take detours on LaGuardia trips to try to make up for any lost earnings that they otherwise could have made on a JFK trip. The median earnings per minute on LaGuardia trips is \$1.46/minute and on JFK trips is \$1.50/minute. As such, the earnings difference is \$1.52 for the median JFK trip, and \$1.00 for the median LaGuardia trip, which is much smaller than the amount gained from a detour, so it is difficult to account for the behavioral response as drivers compensating for 'lost' earnings on LaGuardia trips.

Combined, the results of the analysis on LaGuardia and JFK trips are consistent with increased profitability of detours at LaGuardia inducing drivers to make more LaGuardia trips. I interpret this as evidence that individual differences in propensities to cheat may induce behavioral responses in individuals choosing to place themselves in more cheating situations.

\subsection{Robustness Checks}
In this section, I report the results of several sets of robustness checks on the main analysis.

\subsubsection{Alternative Detour Distance Measures}
One potential concern is that the average detour distance used in the main analysis, which is the in-sample detour distance, only captures the relative differences in propensities to cheat across drivers, rather than the absolute levels of propensities to cheat. This is because the in-sample detour distance by construction has mean 0 across all trips. I address this by running the main analysis using Google Maps-benchmarked detour distances for each trip. \autoref{persistence_gmap_LGALGA_time_LGA1HLGA2H} and \autoref{persistence_gmap_LGAJFK_time_LGA1HJFK2H} present the results of this exercise. The main results hold in this exercise: there is still substantial variation in drivers’ propensities to take detours and that any given driver’s detour behavior is consistent over time. If anything, the correlation is greater and more persistent (correlation = 0.72, t-stat = 94.46). Similarly, no economically significant correlation exists in the JFK placebo test (correlation = 0.08, t-stat = 2.90).

\subsubsection{Random Splits}
I also split the data using a random split instead of a split by time as a robustness exercise. I randomly split the set of trips in the sample into two halves and compute each driver's average detour distance in each half. \autoref{persistence_gmap_LGALGA_rand_LGA1HLGA2H} and \autoref{persistence_insample_LGALGA_rand_LGA1HLGA2H} present the correlation between each driver's average detour distance in each half, as before. The correlations remain large and significant (correlation = 0.78, t-stat = 112.85 for in-sample detour distance; correlation = 0.74, t-stat = 98.02 for Google Maps detour distance).

\subsection{Importance of the heterogeneity}
In this section, I discuss the importance of the heterogeneity of cheating in two ways. First, I compare information on the identity of the driver to information on the time and location of the ride in explaining trip distances. It is possible that even though substantial heterogeneities exist in the level of cheating across drivers, a driver's personal decision to cheat may still be of second order importance in relation to the time and location of the trip or other factors. In contrast to this, I find that who drives the taxi is of primary importance in explaining detours. Second, I calculate the hypothetical additional earnings a driver makes from engaging in this form of cheating and show that the additional earnings can make up to around \$1,767 per year, or 5.2\% of the median New York City taxi driver's annual income.

\subsubsection{Who Drives vs. Where and When?}
I use models with different specifications to compare the explanatory powers of different variables in explaining variations in trip distances. \autoref{variance_decomposition} presents the results. Columns 1 and 2 report the estimated models using the times, dates, and pickup and drop-off locations of the trips. The highest R-squared and adjusted R-squared are 0.31 and 0.15, respectively. Column 3 reports the estimated model using only the identity of the driver. The R-squared and adjusted R-squared are 0.43 and 0.38, respectively, which are significantly higher than the models using time and location predictors.

One concern may be that this is simply capturing variations in the times of when different drivers drive. To test this, I estimate a full model using all the predictor variables. Column 4 reports the full model using time, location, and driver identity data. The R-squared and adjusted R-squared are 0.64 and 0.50, respectively, which are significantly higher than that of the smaller models. Overall, the results suggest that even controlling for the time and locations of the trip, who drives the taxi still has significant power in explaining detouring behavior. Taken individually, who drives the taxicabs seem to be as important as, if not more than, when and where the taxi trip takes place in predicting detours.

\subsubsection{Magnitude of cheating?}
\label{magnitude_of_cheating}
I also show that the earnings from this cheating behavior can comprise a material portion of a driver’s annual income. Consider a driver in the sample who had an average detour distance of 1.957 miles per trip in the sample. The driver made 100 trips from LaGuardia airport to midtown in 2013 and 301 trips from LaGuardia to all of Manhattan in 2013. Assuming that the driver maintains his average detour distance of 1.957 miles per trip on all of his LaGuardia to Manhattan trips means that the driver drove 589 additional miles, which at the standard rate of \$2.50/mile and 20\% tip rate, equates to \$1,767 of additional earnings for the driver in 2013. This amounts to an extra 5.2\% in income for the median New York City taxi driver making \$34,020 per year\footnote{Source: US Bureau of Labor Statistics.}. 

This analysis makes several key assumptions. First, average detour distances may be different between trips to midtown and trips to other parts of Manhattan, although this difference is likely to be small as both types of trips cross the same set of bridges and tunnels into Manhattan, which are the main sources of detours. Second, the average detour distance could also be different if the compositions of passengers are different, which could affect the detour behavior of drivers (see e.g. \cite{balafoutas2013} and \citeauthor{liu2018} (forthcoming)). Third, not all passengers may provide a 20\% tip, so the total tip the driver receives may be lower than 20\%. Finally, there is some opportunity cost of time for the driver to take a longer detour which takes away from the time the driver could have spent picking up other passengers. In light of these assumptions, the calculation presents an estimate of how much additional income a driver who takes detours from LaGuardia airport may earn each year.

%%%%%% Cultural attitudes
\section{Cultural attitudes or learning?}
The previous section analyzed the extent to which cheating behavior varies across individuals and found that there is substantial heterogeneity in propensities to cheat across individuals, and that this heterogeneity induces a behavioral response in the frequency with which individuals place themselves in cheating situations. This section tests different theories of why such heterogeneity exists. Two main hypotheses may explain this heterogeneity. One hypothesis is self-selection, in which individuals have intrinsically different propensities to cheat. These variations in propensities may arise from differences in beliefs or cultural attitudes about cheating (\cite{fisman_miguel_2007}, \cite{barr2010}, and \cite{gächter_schulz_2016}). 
Another hypothesis is learning-by-doing, where drivers learn to cheat through experience over time (\cite{arrow_1962}, \cite{levitt_list_syverson_2013}, \cite{haggag2017learning}, \cite{cook2018}). I find evidence consistent with the self-selection hypothesis and inconsistent with learning-by-doing. I find that cultural attitudes are plausibly related to self-selection into cheating.

\subsection{Cultural attitudes hypothesis}
One reason why individuals have different propensities to cheat may be differences in cultural attitudes. Individuals from countries with different cultural attitudes about cheating may therefore have different beliefs about how acceptable or normalized it is to cheat strangers. These attitudes may then be reflected in differential behaviors in individuals in settings outside their home countries. I present evidence in this section consistent with this hypothesis, that heterogeneity in individual cheating behavior may be related to the cultural attitudes of the individuals’ home countries. The results in this section thus relate to the line of research on the effects of cultural attitudes on economic behavior (\cite{fisman_miguel_2007}, \cite{barr2010}, and \cite{gächter_schulz_2016}). One challenge is that attitudes are inherently hard to define. In the subsequent analysis, I have taken attitudes, broadly defined, to refer to norms and social and intrinsic preferences that lead to different behaviors in the same economic setting. In the present context of detour behaviors in taxi drivers, attitudes may account for why different groups of taxi drivers may make different decisions on whether or not to take detours. 

I test the hypothesis that cultural attitudes are related to differential detouring behaviors in taxi drivers using the matched dataset linking driver trip-level data to the cultural attitudes of the drivers’ home countries. First, I exploit the quasi-random assignment of taxi drivers to trips at each airport to estimate the effect of drivers' home country cultural attitudes on the detour distances of each trip. Since drivers are quasi-randomly assigned to trips at each airport, if cultural attitudes do indeed affect drivers' decisions to take detours, then the amount of detouring on each trip being correlated with the cultural attitudes of the driver's predicted home countries would be consistent with the cultural attitudes hypothesis. This relationship should hold on LaGuardia to Manhattan trips, where longer detours are plausibly cheating. As a placebo test, for flat rate JFK to Manhattan trips, however, there should be no such relationship. As such, I derive two predictions concerning the relationships between the amount of cheating on each trip and the cultural attitudes of the driver:

\begin{itemize}
    \item \textbf{Prediction 1}: drivers from higher corruption countries take longer detours on LaGuardia to Manhattan trips.
    \item \textbf{Prediction 2}: drivers from higher corruption countries do not take longer detours on JFK to Manhattan trips.
\end{itemize}

Second, at the driver level, higher propensities to cheat should also lead to more self-selection into taking cheating trips. As such, cultural attitudes of the drivers' predicted home countries should be correlated with the number of LaGuardia to Manhattan trips. As a placebo test, cultural attitudes should not be correlated with the number of JFK to Manhattan trips. In equilibrium, the higher incentive to take LaGuardia trips relative to JFK trips may actually incentivize drivers to substitute away from making flat rate JFK trips, so it is possible that drivers from higher corruption countries may take fewer JFK trips.

\begin{itemize}
    \item \textbf{Prediction 3}: drivers from higher corruption countries make more LaGuardia trips.
    \item \textbf{Prediction 4}: drivers from higher countries countries do not make more JFK trips.
\end{itemize}

For predictions 1 and 2, the empirical analysis I run is as follows:

$$Y_{i} = \alpha_{0} + \alpha_{1} \cdot X_{i} + \alpha_{2} \cdot Z_{i} + \epsilon_{i}$$

Where $Y_{i}$ is the trip distance on LaGuardia trip i for prediction 1 and on JFK trip i for prediction 2. $X_{i}$ is the average 2013 CPI score of the name-origin group of countries that the driver is predicted to be from\footnote{See \hyperref[appendix_culture_data]{Appendix A} for a description of the name prediction software.}, multiplied by -1/10 and weighted by the total number of drivers from each country. As such, a higher CPI score is associated with countries perceived to be more corrupt. $Z_{i}$ is a set of trip-level time, destination, and Google Maps controls.

For predictions 3 and 4, the empirical analysis I run is as follows:

$$Y_{i} = \alpha_{0} + \alpha_{1} \cdot X_{i} + \epsilon_{i}$$

Where $Y_{i}$ is the number of LaGuardia trips driver i made in 2013 for prediction 3 and the number of JFK trips driver i made in 2013 for prediction 4. $X_{i}$ is the average 2013 CPI score of the name-origin group of countries that the driver is predicted to be from, multiplied by -1/10 and weighted by the total number of drivers from each country. As such, a higher CPI score is associated with countries perceived to be more corrupt.  

\subsection{Results}
\autoref{cultural_regs_severity} reports the regression results of predictions 1 and 2. \autoref{ethnicity_regs_LGA_cpi2013weighted_ethnicitylevel_min100} and \autoref{ethnicity_regs_JFK_cpi2013weighted_ethnicitylevel_min100} plot the results at the name-origin country-group level for name-origin country-groups with at least 100 drivers. Drivers from higher-corruption countries take longer detours on LaGuardia trips on average. An increase in the constructed CPI score of 1 (corresponding to a decrease of 10 in Transparency International's CPI scores) is associated with an increase in average driver detour distance of 0.0395 (Google Maps) to 0.0420 (In-Sample) miles. The coefficients are statistically distinguishable from zero at the 1\% level. In the placebo test, higher corruption in home countries are not correlated with greater detour distances on JFK trips. As such, the empirical relationship between cultural attitudes are consistent with predictions 1 and 2.

\autoref{cultural_regs_frequency} reports the regression results of predictions 3 and 4. Drivers from higher-corruption countries make more LaGuardia trips and make fewer JFK trips. An increase in the constructed CPI score of 1 (corresponding to a decrease of 10 in the original CPI scores) is associated with 0.295 more LaGuardia trips per year and 0.430 fewer JFK trips per year. Standard errors are clustered at the name-origin group level. The coefficients are statistically distinguishable from zero at the 1\% level\footnote{Alternative specifications of computing standard errors, including \cite{cameron2008bootstrap} robust cluster size corrections, are reported in \hyperref[cultural_regs_severity_clusterse]{Table A1}. The results remain robust under alternative specifications.}. The positive coefficient on CPI Score in the LaGuardia regression is consistent with prediction 3 in that a higher corruption norm in the predicted home countries is correlated with more LaGuardia trips made by the driver. The negative coefficient on the CPI Score in the JFK regression is consistent with prediction 4. The combination of positive coefficient for LaGuardia trips and negative coefficient for JFK trips on CPI scores highlights a plausible substitution effect from fewer flat-rate JFK trips to more LaGuardia trips by drivers predicted to be from higher-corruption norm countries.

To understand the magnitude of the relationship, an increase in the corruption perception index score of 50 points (for example, between Nicaragua (28) and Germany (78)) is associated with a decrease in the average detour distance on LaGuardia to Manhattan trips of about 0.19 (Google Maps) to 0.21 (in-sample) miles, which equates to 45\%-50\% of the average detour distance across all LaGuardia to Manhattan trips. The same 50-point increase in the corruption perception index score would also be associated with 1.5 fewer LaGuardia trips and 2.1 more JFK trips per driver per year.

Taken together, the results on the detour distances (predictions 1 and 2) and the results on the number of trips of a specific airport (predictions 3 and 4) are consistent with the four predictions of the cultural attitudes hypothesis. The findings are plausibly consistent with cultural attitudes as an origin of heterogeneity in cheating.

\subsection{Robustness}
\subsubsection{Results Robust to Inclusion/Exclusion of Specific Countries?}
One concern is that the results may be disproportionately driven by drivers from specific countries. In particular, French name origins include a range of countries for which the name classification algorithm cannot distinguish. I address this issue by re-running the analysis and removing drivers from one specific name-origin country-group each time. \autoref{ethnicity_removeone_regs} reports the results of the robustness exercise. The results remain qualitatively similar to the full-sample estimates and are consistent with the predictions of the cultural attitudes hypothesis. Specifically, the coefficients on the regression excluding French name origins remain economically and statistically significant, and is larger in magnitude than the coefficient estimated on the full sample, which is suggests that the results are more robust on the subsample where name-origin group links are more certain.

\subsubsection{Cultural Attitudes Scores Weighting}
Another concern is that the results may be sensitive to the weighting of cultural attitudes scores by likely home countries. In the main analysis, I use a weighted average of CPI scores for each name-origin country-group, weighted by the number of drivers from each country. I re-run the analysis using an unweighted average of CPI scores instead of a weighted average. \autoref{cultural_regs_severity_unweighted} presents the results of this robustness exercise on predictions 1 and 2. \autoref{cultural_regs_frequency_unweighted} presents the results of this robustness exercise on predictions 3 and 4. The results are robust to the reweighting and remain consistent with the predictions of the cultural attitudes hypothesis.

\subsubsection{CPI Scores from Different Years?}
I also use CPI scores from 1998, the first year in which the CPI scores were broadly available for the set of countries in the sample. It is plausible that since most drivers are foreign-born, corruption norms in home countries in 2013 may not reflect the norms that were prevalent during the times when the drivers were actually present in the home countries. \autoref{cultural_regs_severity_1998} reports the results of this robustness exercise for the detour distances and \autoref{cultural_regs_frequency_1998} reports the results of this robustness exercise for the number of trips. There are fewer trips and drivers in the sample because fewer countries have Corruption Perceptions Index scores available in 1998 than in 2013. The results are robust to using 1998 CPI scores and are similar to using 2013 CPI scores.

\subsubsection{Mapping to Home Country Attitudes}
An additional concern is that there could be measurement noise introduced through the name-to-home country mapping. In particular, names may not be mapped to the correct attitudes of home countries of origin. In the baseline specification, I map individual names to the name-origin group with the highest probability in the NamePrism classification tree (which I will refer to as the "one-to-one" mapping). The concern is that this mapping may be noisy and imprecise for names with multiple likely name-origin groups. In an alternative set of specifications, I address this by using a probability-weighted mapping, where each driver’s attitudes are calculated according to the weights equal to the probability that the individual driver is of each given name-origin group. The probabilities I use are the empirical probabilities from the NamePrism training dataset. For example, a driver whose name has an empirical probability of 99\% of being of origin country A and 1\% of being of origin country B, then his/her assigned norm using probability-weighted mapping will be 0.99*(Norm of Country A) + 0.01*(Norm of Country B), whereas another driver with probabilities 75\% of origin country A and 25\% of origin country B will have an assigned norm of 0.75*(Norm of Country A) + 0.25*(Norm of Country B). In contrast, the original maximum-probability mapping method would have assigned 1*(Norm of Country A) to both of these cases. As such, the probability-weighted mapping will lead to individual-specific corruption index scores, where drivers whose names have higher probabilities of being from a certain country will have scores closer to that country’s than another driver whose name is relatively less likely to be from that country.

I report the results using probability-weighted mapping in \autoref{cultural_regs_severity_proportional}. Columns 1-3 use the original maximum probability (one-to-one) for the mapping, whereas columns 4-6 use the probability-weighted mapping. Additionally, I also report the different results using different cultural attitudes weighting methods: columns 1 and 4 use the simple average across home countries; columns 2 and 5 use the median across home countries; columns 3 and 6 use the weighted average weighted by the number of drivers from each country. There are slightly fewer observations in this sample since to compute the weighted averages, I used the TLC administrative reports which did not report countries for which fewer than 10 drivers were from, so drivers from those countries were removed from this sample.

The estimated coefficients on the relationship between cultural attitudes and detour behavior are positive and statistically significant in all specifications. Using more robust mapping methods (i.e. using probability-weighted mapping to set of possible home countries) produce significantly larger magnitudes in the coefficient estimates than the one-to-one mappings. The point estimate using probability-weighted mappings (columns 4-6) are each larger than their respective maximum-probability mapping (columns 1-3) counterparts. 

Similarly, using more robust cultural attitudes score weightings (i.e. taking weighted average as opposed to simple average or median) also results in coefficients significantly larger in magnitude. Overall, it seems that the relationship between cultural attitudes and detour behavior is positive, significant, and fairly robust to different mapping methods. If anything, using more robust mapping methods seems to lead to larger point estimates.

\subsubsection{Demands for money}
An alternative explanation is that cultural attitudes are correlated with cheating because drivers from different home countries have different demands for money. To test for the effects of demands for money, I exploit differential variations in demand for money in the time periods around significant cultural and faith dates: Christmas\footnote{Christmas: US Census, Facts for Features, Special Edition.}, Lunar New Year\footnote{Lunar New Year: CNBC.}, and Ramadan\footnote{Ramadan: BBC News.}. Each of these significant cultural and faith dates has well-documented associated increases in consumer spendings.

The empirical strategy is to compute the average detour distances of all drivers of a respective name-origin group who are plausibly affected by the cultural and faith date and identify any differential behaviors by these drivers during this time period. I calculate the abnormal detour distance, defined as the difference between the average detour distance of drivers of the affected name-origin group and of unaffected name-origin groups. If drivers do detour in response to differential marginal utilities, then the affected drivers' average detour distances should increase relative to other drivers\footnote{Christmas was December 25, 2013. Lunar New Year was February 10, 2013. Ramadan was July 8 to August 7, 2013.}.

For Christmas, I define drivers from home countries with at least 80\% of the population identifying as Christians as the affected drivers, and drivers from home countries with less than 20\% of the population identifying as Christians as the unaffected drivers. For Lunar New Year, I define drivers from East Asian countries as the affected drivers and all other drivers as non-affected drivers. For Ramadan, I define drivers from home countries with at least 80\% of the population identifying as Muslims as the affected drivers, and drivers from countries with less than 20\% of the population identifying as Muslims as the unaffected drivers.

\autoref{marginalutility_christians} presents the analysis around Christmas. \autoref{marginalutility_cny} presents the analysis around Lunar New Year. \autoref{marginalutility_muslims} present the analysis around Ramadan. The red dashed lines indicate the significant cultural and faith dates (for Ramadan, the two lines indicate the beginning and end of Ramadan). The horizontal axis represents the week number and each point represents the abnormal detour distance of the affected group during that week. For the analysis of Christmas, I omit week 53 because it is a single day rather than a full week. A greater abnormal detour distance represents longer detours taken by the affected group relative to the unaffected group. No significant differentiation in detouring behavior is observable in the weeks before, during, and after the dates. This is inconsistent with drivers exhibiting differential propensities to cheat due to variations in demand of money over time.

\subsection{Learning-by-doing hypothesis}
The alternative hypothesis to cultural attitudes-based self-selection is learning-by-doing (\cite{arrow_1962}, \cite{levitt_list_syverson_2013}, \cite{haggag2017learning}, \cite{cook2018}). In learning-by-doing, drivers learn to take detours through experience over time. I test this hypothesis by analyzing the detour behaviors of drivers as they gain experience over time. The empirical specification is as follows:

$$Y_{i} = \alpha_{0} + \alpha_{1} \cdot X_{i} + \alpha_{2} \cdot Z_{i} + \epsilon_{i}$$

Where $Y_{i}$ is a measure of the trip distance, $X_{i}$ is the number of previous trips a driver has made, which I use as a proxy for experience, and $Z_{i}$ is a fixed effect indicator for the driver. Thus, the coefficient of interest is $X_{i}$, which captures the relationship between the detour distance on a given trip and the number of trips driven within a particular individual. If drivers do indeed learn over the course of their trips, the coefficient of interest, $\alpha_{1}$, should be positive and statistically significant.

\autoref{learning_regs_fullsample} reports the results of the analysis. In the three specifications, the coefficients of interest $\alpha_{i}$ are negative and statistically indistinguishable from zero. The point estimates are also not economically meaningful.

I augment this analysis by considering the subpopulation for whom the effects of learning, if they exist, would plausibly be the strongest: new drivers. New drivers plausibly would experience the most learning-by-doing, as compared to drivers with more experience, because drivers may be more likely to incrementally learn certain driving and detouring behaviors after their first few trips than after already having made many trips. I thus repeat the analysis and restrict the sample to trips made by only new drivers, here defined as drivers who made their first trip on or after June 1, 2013, which I use as a proxy for being a new driver. \autoref{learning_regs} presents the results of this analysis. In the sample of new drivers, there is no statistically or economically significant relationship between the detour distance on each trip and how many previous trips the driver has made, controlling for the identity of the driver.

There are a few potential concerns to this analysis. One potential concern with the new driver analysis is that the proxy for new drivers may also include drivers who are not new but rather simply did not drive for an extended period of time from the beginning of 2013 until after June 1, 2013. To address this, I repeat the analysis by using different samples with different cutoff dates, for example, restricting the sample to only drivers who had their first trip after September 1, 2013. The Post-March 1 and Post-September 1 columns of \autoref{learning_regs} present the results of this robustness exercise. No statistically or economically significant relationship between detour behavior and experience exists in the alternate samples. As such, the evidence appears inconsistent with the learning hypothesis in explaining variations in cheating across individuals and consistent with the cultural attitudes-based hypothesis.

Furthermore, the sample of new drivers I consider is necessarily a smaller subset of all drivers. It is possible that the results in the new drivers analysis could be different if it were a different set of drivers in the dataset. This is indeed a limitation to the new drivers analysis, and is one potential direction of future work in using either experimental settings or other large samples of new individuals to study learning effects in cheating.

Another potential concern is that the rate of learning may take a form different than the linear learning curve implied here. \autoref{learning_plot_fullsample_1} plots the trip distances on trips by the number of all previous taxi trips made, residualized by driver fixed effects for the full sample of drivers. \autoref{learning_plot_newdrivers_1} plots the analogous plot for the subset of new drivers. I find that there does not appear to be a non-linear learning curve. In spite of this, it is still possible that drivers may learn at different rates or learn before they begin their first trip, the latter of which I cannot observe in this dataset.

%%%%% Conclusion
\section{Conclusion}
This paper studies the behaviors and causes of cheating at the individual level using a novel methodology of detecting cheating by taxi drivers. I document three main findings. First, there is significant detouring on average by drivers. These detours do not save time or avoid traffic, and drivers do not detour when there is no financial incentive to do so. Second, there is significant heterogeneity in cheating across individuals, yet each individual's propensity to cheat is stable: drivers who detour almost always detour, while those who do not detour almost never do. Drivers who take longer detours on each trip also take such cheating trips more often. Third, cultural attitudes plausibly explain some of this heterogeneity in cheating across individuals.

The findings of this paper empirically join the two strands of literatures on cheating and on cultural attitudes by documenting individual heterogeneity in cheating and linking it to origins based in cultural attitudes. Given the limited amount of demographic information on individuals available in this setting, further analysis could benefit from considering settings or collecting data where more detailed demographic information would be available, for instance on the length of time the individuals have been in the U.S. and factors such as education levels and income. Applying similar analysis to taxi trips in other locations could also be helpful in understanding how the findings in New York City may generalize to other cities. Extending the questions raised in this paper, how attitudes affect economic behavior in other domains and how peer networks may affect individuals’ propensities to engage in certain types of economic behaviors remain open questions and directions of future work.

\end{doublespacing}

%%%%% Bibliography
\newpage
\bibliographystyle{apalike}
\bibliography{mybib}
%\printbibliography[mybib.bib]
\label{references}
\nocite{*}

%%%%% Plots
\newpage
\section{Tables and Figures}

%%% Tables

% Tables 2-4: Summary Stats
\textbf{Number of Rides:}
\begin{table}[H]
\centering
\caption{Number of Rides}
\label{summary-stats}
\begin{tabular}{ll}
\hline
Total Number of Rides in 2013   & 121,964,746 \\ \hline
Total LGA-to-Midtown East Rides & 296,750     \\
Total JFK-to-Midtown East Rides & 137,472     \\ \hline
\end{tabular}
\end{table}

\textbf{Trip Statistics: LaGuardia-to-Midtown Manhattan Rides}
\begin{table}[H]
\centering
\caption{Trip Statistics: LaGuardia-to-Midtown Manhattan Rides}
\label{my-label}
\begin{tabular}{ccccccc}
                      & Minimum & 1Q    & Median & Mean  & 3Q     & Max    \\ \hline
Trip Distance (miles) & 6.030   & 8.820 & 10.100 & 9.881 & 10.800 & 19.940 \\ 
Trip Time (minutes)   & 1       & 19.80 & 25.00  & 26.83 & 32.00  & 317.60 \\ 
Fare Amount (USD)     & 2.50    & 27.50 & 30.50  & 30.59 & 33.50  & 128.00 \\ 
Tip Amount (USD)      & 0       & 0     & 6      & 4.826 & 7.550  & 81.02  \\ \hline
\end{tabular}
\end{table}

\textbf{Trip Statistics: JFK-to-Midtown Manhattan Rides}
\begin{table}[H]
\centering
\caption{Trip Statistics: JFK-to-Midtown Manhattan Rides}
\label{my-label}
\begin{tabular}{ccccccc}

                      & Minimum & 1Q    & Median & Mean  & 3Q    & Max    \\ \hline
Trip Distance (miles) & 9.70    & 17.00 & 17.40  & 17.59 & 17.83 & 35.01  \\
Trip Time (minutes)   & 1       & 29.00 & 38.00  & 40.34 & 48.37 & 750    \\
Fare Amount (USD)     & 2.50    & 52    & 52     & 50.98 & 52    & 310    \\
Tip Amount (USD)      & 0       & 0     & 5      & 5.641 & 11.45 & 116.20 \\ \hline
\end{tabular}
\end{table}

The tables provide summary statistics of the New York City taxi trip-level dataset. The first table reports the number of yellow taxicab rides in 2013 that took place between each of the destinations. The second table reports summary statistics of trips from LaGuardia airport to midtown Manhattan. The third table reports summary statistics of trips from JFK airport to midtown Manhattan. In the latter two tables, the fare amount is the portion of the total cost of the ride that is the base or fixed fare plus any distance or time-based components. The fare amount does not include any surcharges or tolls or tips. 2.7\% of JFK trips do not have a fare amount of \$52 because the driver did not set the meter to the JFK fare code. I chose to keep these trips in-sample because they still are taxi trips made from JFK airport to Manhattan.

% Table 1: Detour Distances
\newpage
% \usepackage{multirow}
\begin{table}[H]
\centering
\caption{Detour Distances by Airport}
\label{detour-tables}
\begin{tabular}{cccccc}
                                                                                                  &                   & LaGuardia & JFK   & Difference & t-stat       \\ \hline
\multicolumn{6}{c}{Detour Distance Percentages}                                                                                                                     \\
                                                                                                  & Average           & 5.6\%     & 1.5\% & 4.1\%      & {[}107.94{]} \\ 
\multirow{3}{*}{\begin{tabular}[c]{@{}c@{}}Percent of Trips with\\ Detour Distances\end{tabular}} & 10\% or more      & 32.4\%    & 8.3\% & 24.1\%     & {[}212.42{]} \\ 
                                                                                                  & 20\% or more      & 17.5\%    & 4.9\% & 21.6\%     & {[}138.96{]} \\ 
                                                                                                  & 30\% or more      & 8.0\%     & 0.7\% & 7.3\%      & {[}133.41{]} \\ \hline
\multicolumn{6}{c}{Detour Distance Miles}                                                                                                                           \\ 
                                                                                                  & Average           & 0.42      & 0.23  & 0.19       & {[}39.25{]}  \\
\multirow{3}{*}{\begin{tabular}[c]{@{}c@{}}Percent of Trips with\\ Detour Distances\end{tabular}} & 1 mile or more    & 32\%      & 11\%  & 21\%       & {[}180.8{]}  \\ 
                                                                                                  & 1.5 miles or more & 24\%      & 9\%   & 15\%       & {[}143.21{]} \\ 
                                                                                                  & 2 miles or more   & 14\%      & 8\%   & 6\%        & {[}65.15{]}  \\ \hline
\end{tabular}
\end{table}

This table presents the detour distances of all taxi trips from LaGuardia airport to midtown Manhattan and trips from JFK airport to midtown Manhattan in 2013. The detour distances are calculated by comparing the actual trip distance (in miles) to the optimal Google Maps trip distance for the same trip at the exact same time of day and day of week. The top panel presents the detour distances as a percentage of the optimal Google maps trip distance and the bottom panel presents the detour distances in miles. The t-stat is on a two-sample t-test of difference of means between the LaGuardia trips and the JFK trips.

% Table 7: LGA Time Regressions
\newpage
\begin{table}[H]
\centering
\caption{LaGuardia Time Regressions}
\label{lga-time-regressions}
\begin{tabular}{cccc}
\hline
                                                                & (1)       & (2)                                                                      & (3)                                                                    \\ 
\begin{tabular}[c]{@{}c@{}}Dependent\\   variable:\end{tabular} & Trip Time & \begin{tabular}[c]{@{}c@{}}Time Discrepancy Google\\   Maps\end{tabular} & \begin{tabular}[c]{@{}c@{}}Time Discrepancy\\   In-Sample\end{tabular} \\ 
                                                                &           &                                                                          &                                                                        \\ 
\begin{tabular}[c]{@{}c@{}}Detour\\   Distance\end{tabular}     & 0.382***  & 0.436***                                                                 & 0.376***                                                               \\ 
                                                                & (0.0187)  & (0.0184)                                                                 & (0.0188)                                                               \\ 
Constant                                                        & 26.67***  & 2.858***                                                                 & -0.159***                                                              \\ 
                                                                & (0.0221)  & (0.0213)                                                                 & (0.0222)                                                               \\ 
                                                                &           &                                                                          &                                                                        \\ 
\begin{tabular}[c]{@{}c@{}}Time FEs\end{tabular}            & Yes       & Yes                                                                      & Yes                                                                    \\ 
\begin{tabular}[c]{@{}c@{}}Destination FEs\end{tabular}     & Yes       & Yes                                                                      & Yes                                                                    \\ 
                                                                &           &                                                                          &                                                                        \\ 
Observations                                                    & 296,748   & 296,748                                                                  & 294,378                                                                \\ 
R-squared                                                       & 0.403     & 0.340                                                                    & 0.004                                                                  \\ \hline
\multicolumn{4}{l}{\begin{tabular}[c]{@{}l@{}}Standard   errors in parentheses\end{tabular}}                                                                                                                                \\ 
\multicolumn{4}{l}{\begin{tabular}[c]{@{}l@{}}***   p\textless0.01, ** p\textless0.05, * p\textless0.1\end{tabular}}                                                                                                        \\ 
\end{tabular}
\end{table}

This table reports the results of the regressions of trip time on trip distance on LaGuardia to Manhattan taxi trips. In the regressions, the dependent variable is one of three measures of time, all in minutes: (1) recorded trip time, (2) time discrepancy from the trip time predicted by Google Maps, and (3) time discrepancy compared to the average trip time of all other similar taxi trips in the sample. The independent variable is the detour distance in miles. A positive time discrepancy corresponds to a longer trip time in column (1), a longer trip time than the Google Maps predicted time in column (2), and a longer trip time than the average trip time of all other trips happening within the same hour on the exact same date going from LaGuardia airport to the same block in Manhattan.

% Table 8: LGA Risk Aversion
%\newpage
%\input{plots/lga-variance.tex}

% Table 9: LGA Left Tail
%\newpage
%\input{plots/lga-left-tail.tex}

% Table 10: Yankee Stadium
\newpage
\begin{table}[H]
\centering
\caption{Yankee Stadium and Residential Area Trips}
\label{yankee-stadium}
\begin{tabular}{ccc}
\hline
                                                                                                & Yankee Stadium & Residential Area   \\ \hline
Avg. Detour Distance                                                                            & -0.17          & -2.209        \\
{[}t-stat{]}                                                                                    & {[}-4.7529{]}  & {[}-22.091{]} \\ 
\begin{tabular}[c]{@{}c@{}}Percentage of trips with\\   greater than 1-mile detour\end{tabular} & 9.60\%         & 5.60\%        \\ 
\begin{tabular}[c]{@{}c@{}}Percentage of trips with\\   greater than 2-mile detour\end{tabular} & 4.20\%         & 3.90\%        \\ \hline
\end{tabular}
\end{table}

\begin{figure}[H]
    \centering
    \caption{Map of Study Areas}
    \includegraphics[height=340pt]{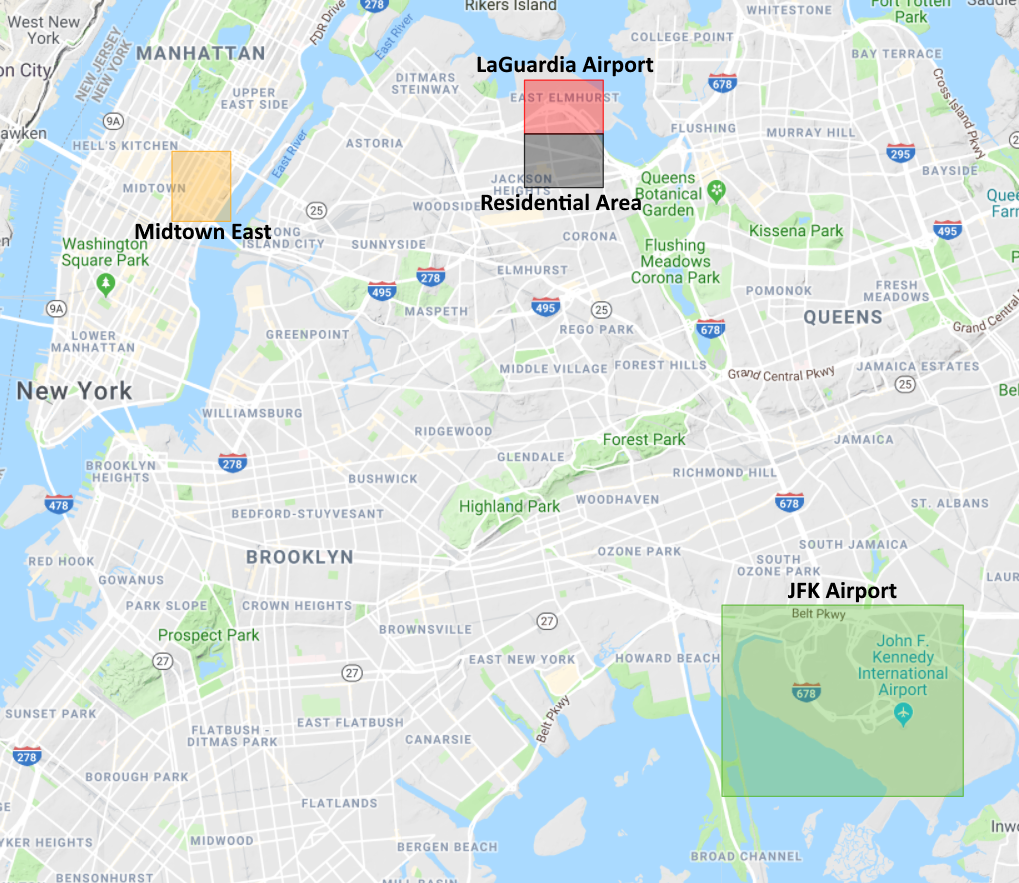}
    \label{fig:map}
\end{figure}

\begin{figure}[H]
    \centering
    \caption{LaGuardia Detour Distances Histogram}
    \includegraphics[height=340pt]{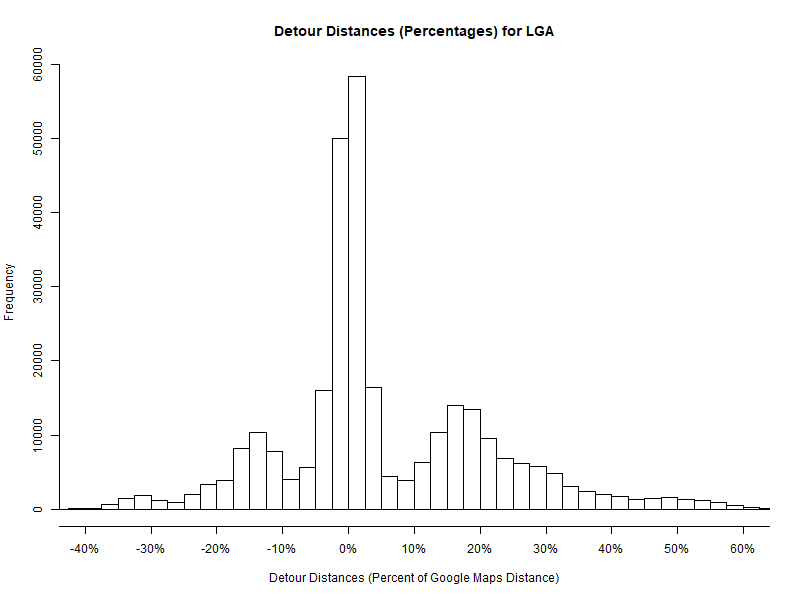}
    \label{detour_histogram_LGA_pct}
\end{figure}

\begin{figure}[H]
    \centering
    \caption{JFK Detour Distances Histogram}
    \includegraphics[height=340pt]{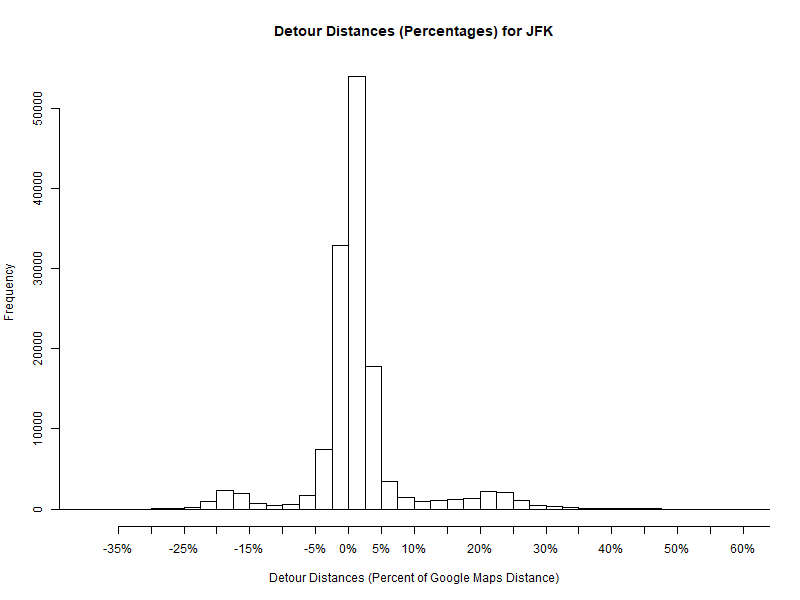}
    \label{detour_histogram_JFK_pct}
\end{figure}

\begin{figure}[H]
    \centering
    \caption{Persistence of Detour Distances (In-Sample)}
    \includegraphics[height=280pt]{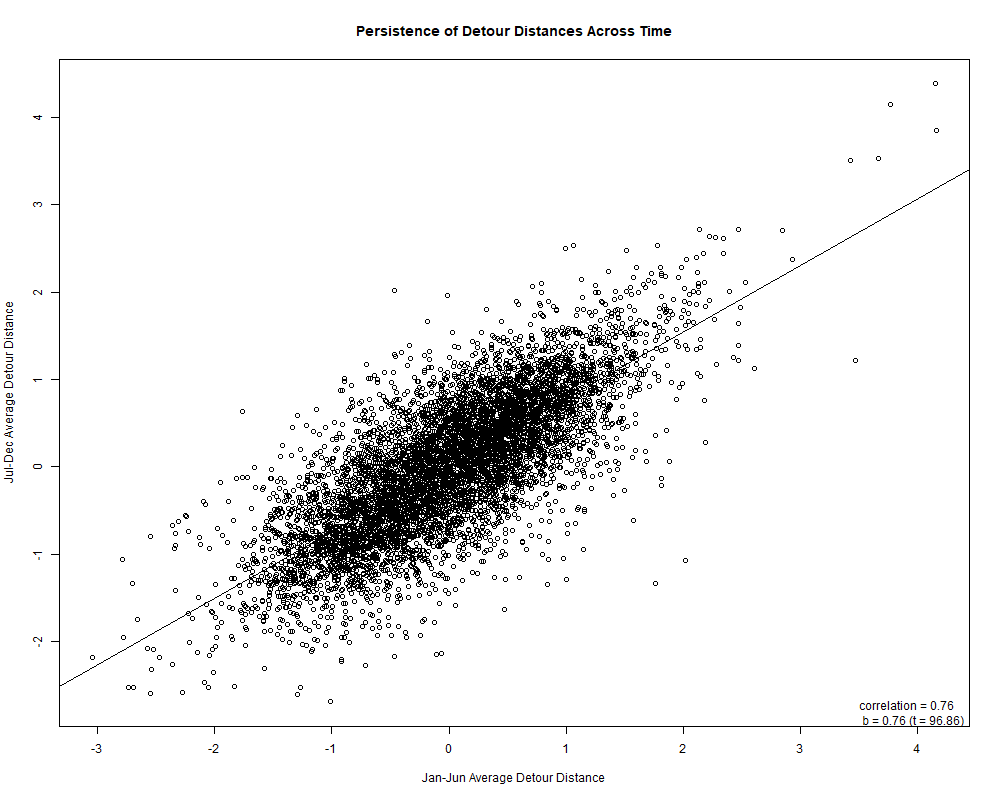}
    \label{persistence_insample_LGALGA_time_LGA1HLGA2H}
\end{figure}

\begin{figure}[H]
    \centering
    \caption{Persistence of Detour Distances (In-Sample): JFK Placebo Test}
    \includegraphics[height=280pt]{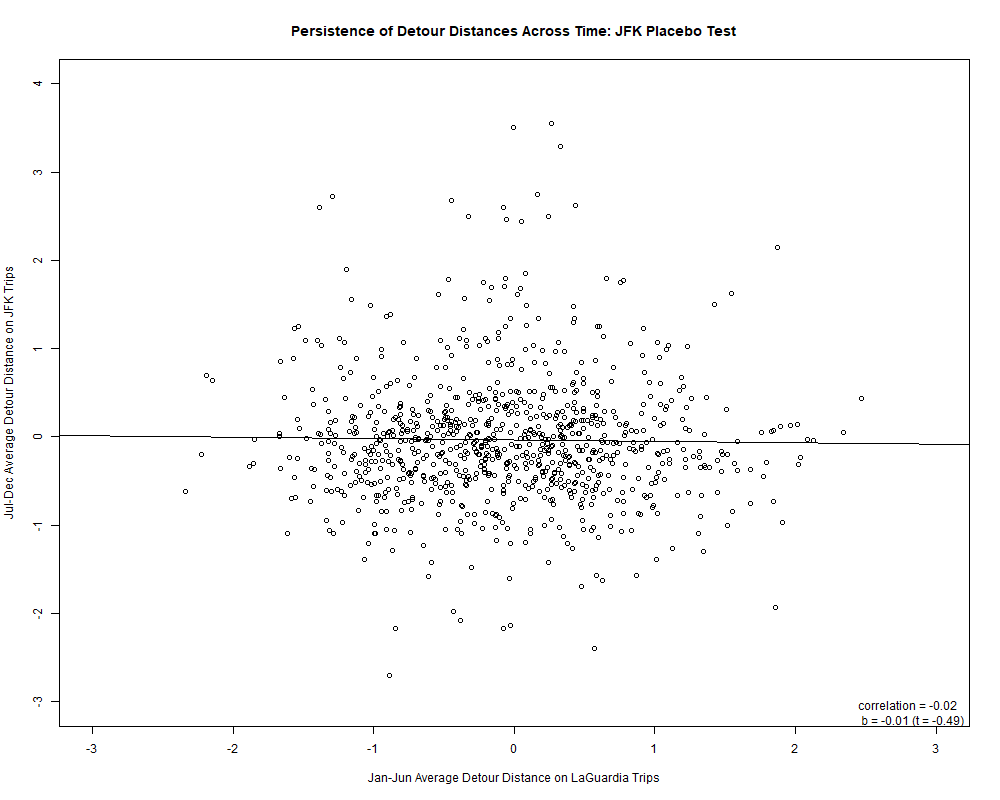}
    \label{persistence_insample_LGAJFK_time_LGA1HJFK2H}
\end{figure}

\begin{figure}[H]
    \centering
    \caption{Persistence of Detour Distances (Google Maps)}
    \includegraphics[height=280pt]{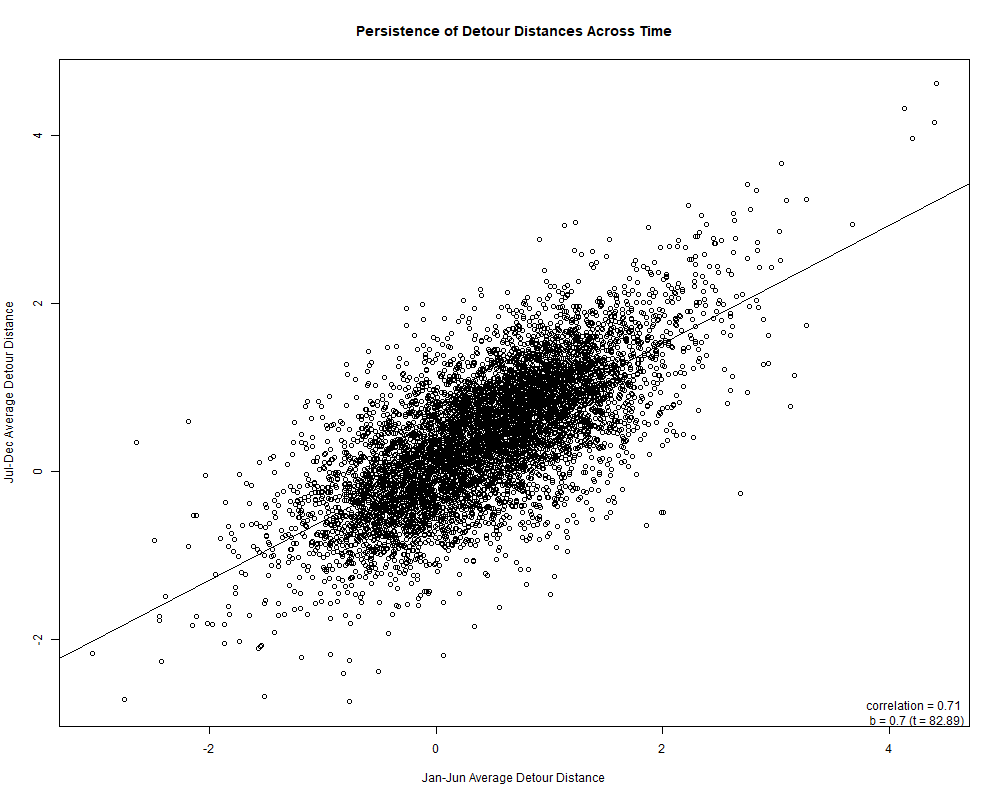}
    \label{persistence_gmap_LGALGA_time_LGA1HLGA2H}
\end{figure}

\begin{figure}[H]
    \centering
    \caption{Persistence of Detour Distances (Google Maps): JFK Placebo Test}
    \includegraphics[height=280pt]{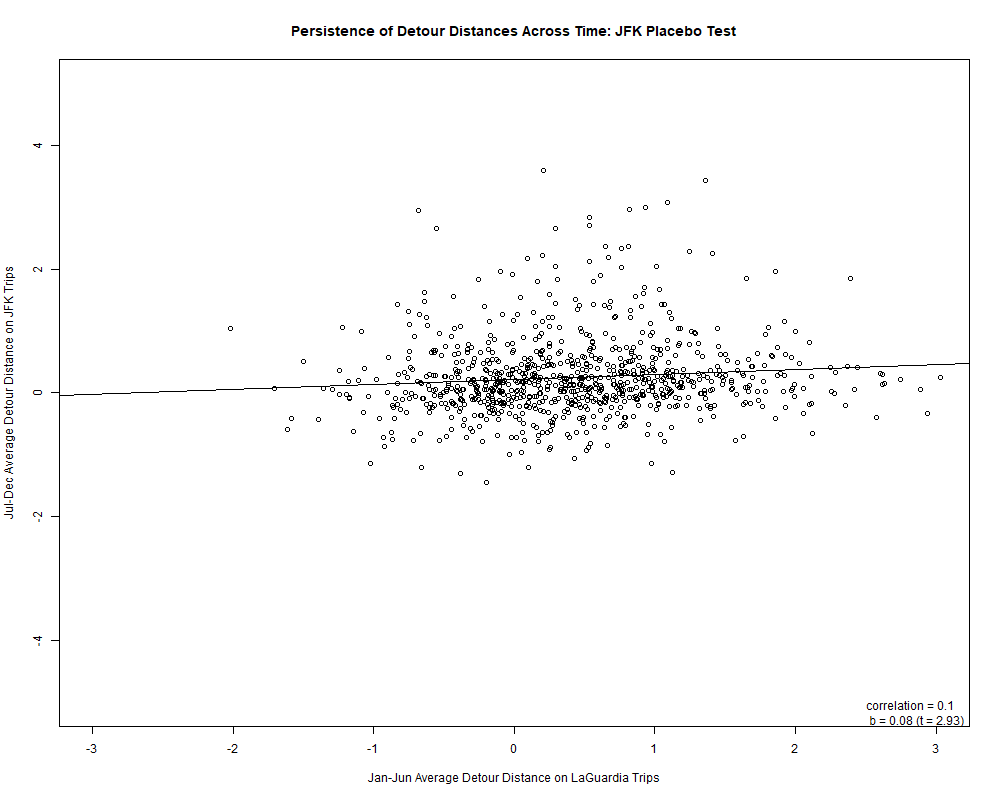}
    \label{persistence_gmap_LGAJFK_time_LGA1HJFK2H}
\end{figure}

\begin{figure}[H]
    \centering
    \caption{Behavioral Response at LaGuardia}
    \includegraphics[height=280pt]{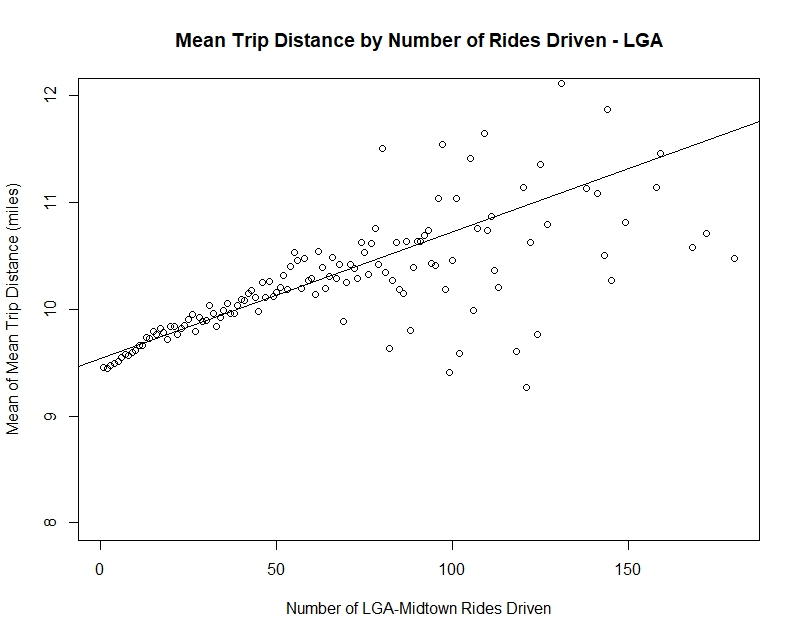}
    \label{LGA_distance_numtrips}
\end{figure}

\begin{figure}[H]
    \centering
    \caption{Behavioral Response at JFK}
    \includegraphics[height=280pt]{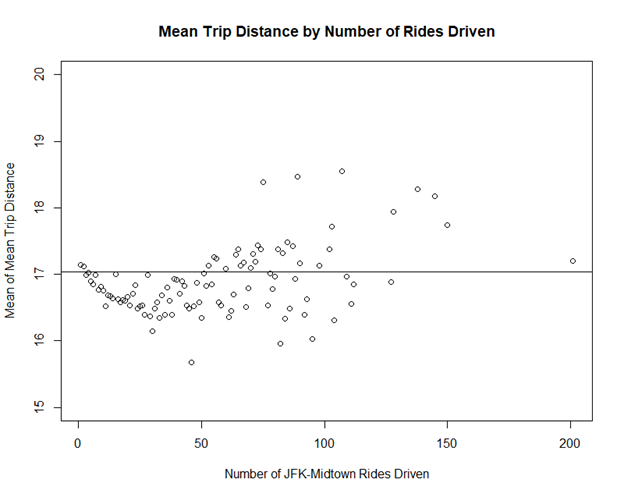}
    \label{JFK_distance_numtrips}
\end{figure}

\begin{figure}[H]
    \centering
    \caption{Detour Distances and Cultural Attitudes by Name-Origin Groups}
    \includegraphics[height=280pt]{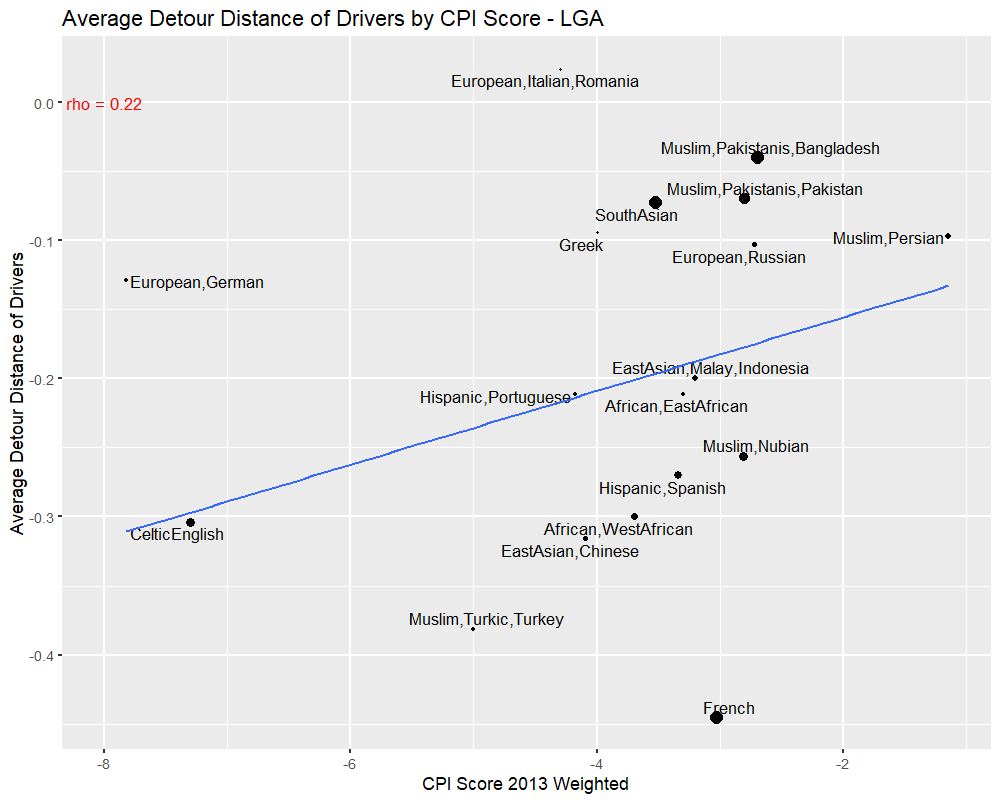}
    \label{ethnicity_regs_LGA_cpi2013weighted_ethnicitylevel_min100}
\end{figure}

\begin{figure}[H]
    \centering
    \caption{Detour Distances and Cultural attitudes by Name-Origin Groups: JFK Placebo Test}
    \includegraphics[height=280pt]{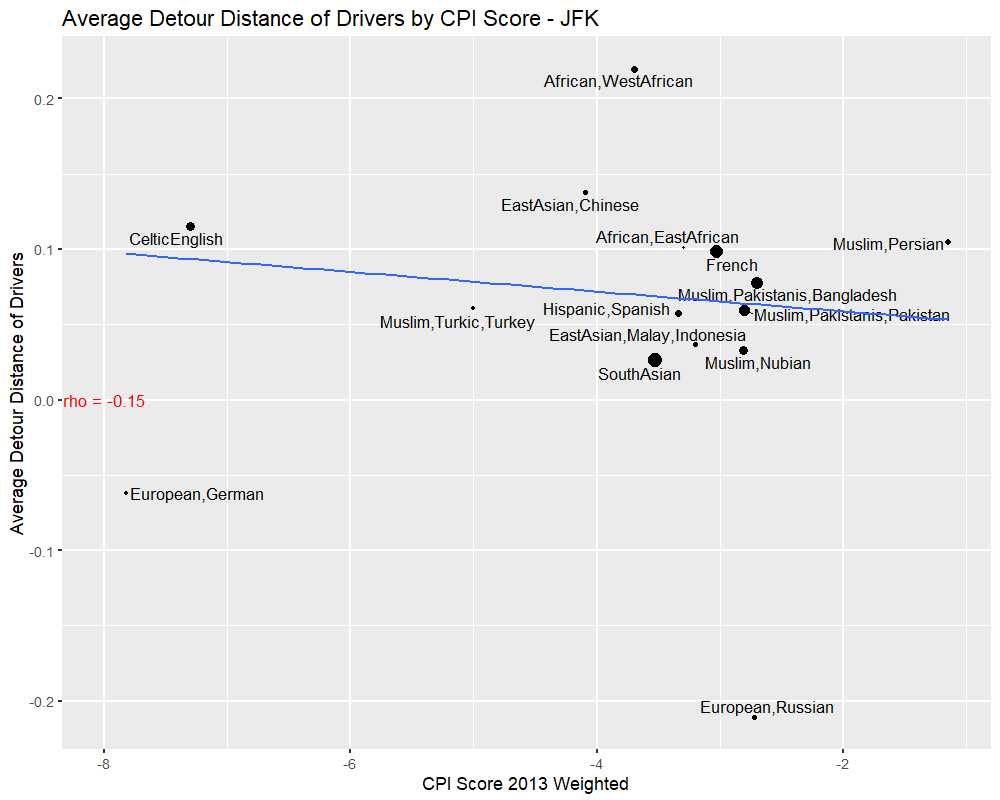}
    \label{ethnicity_regs_JFK_cpi2013weighted_ethnicitylevel_min100}
\end{figure}

This table reports the average detour distances and the percentages of trips with greater than 1- or 2-mile detours for trips from Yankee Stadium or from the Residential Area to Manhattan. The Residential Area is defined as a residential area adjacent to LaGuardia airport in Queens, NY (see: \autoref{fig:map}). The t-statistic of a one-sample t-test with null hypothesis of mean 0 is reported in brackets.

% Table 11: Explanatory Powers
\newpage
\begin{table}[H]
\caption{Who Drives vs. Where and When?}
\label{variance_decomposition}
\begin{tabular}{lcccc}
                                                             & (1)                  & (2)                  & (3)                  & (4)                  \\\hline
VARIABLES                                                    & \multicolumn{4}{c}{Trip Distance}                                                         \\\hline
Fixed Effects (Interacted):                                  &                      &                      &                      &                      \\
Hour of Day                                                  & Yes                  & Yes                  &                      & Yes                  \\
Day of Week                                                  & Yes                  &                      &                      &                      \\
Exact Date                                                   &                      & Yes                  &                      & Yes                  \\
Dropoff Location                                             & Yes                  & Yes                  &                      & Yes                  \\
Driver ID                                                    &                      &                      & Yes                  & Yes                  \\
                                                             & \multicolumn{1}{l}{} & \multicolumn{1}{l}{} & \multicolumn{1}{l}{} & \multicolumn{1}{l}{} \\
Observations                                                 & 296,573              & 275,466              & 292,717              & 270,756              \\
R-squared                                                    & 0.152                & 0.316                & 0.428                & 0.640                \\
Adj. R-squared                                               & 0.146                & 0.154                & 0.379                & 0.503                \\
Root MSE                                                     & 1.189                & 1.177                & 1.013                & 0.901                \\\hline

\end{tabular}
\end{table}

This table reports the explanatory powers of various specifications in explaining the variations in the distance of LaGuardia to Manhattan trips. Fixed effects for hour of day, day of week, the exact date, and the dropoff location are interacted wherever they are included in specification.

% Table 12: Frequency Regs LGA
\newpage
\begin{table}[H]
\caption{Frequency of Visits vs. Severity of Detours}
\label{frequency_regs}
\begin{tabular}{lcccc}
                           & \multicolumn{2}{c}{LGA Detour Distance} & \multicolumn{2}{c}{JFK Detour Distance}           \\\hline
VARIABLES                  & Google Maps         & In-Sample           & Google Maps & In-Sample  \\
                           &                     &                     &             &            \\
Number of LGA Trips Driven & 0.0103***           & 0.0106***           &             &            \\
                           & (0.000584)          & (0.000577)          &             &            \\
Number of JFK Trips Driven &                     &                     & 0.000130    & 0.000571   \\
                           &                     &                     & (0.000455)  & (0.000469) \\
Constant                   & 9.586***            & 0.114***            & 17.59***    & 0.218***   \\
                           & (0.0139)            & (0.0135)            & (0.00998)   & (0.01000)  \\
Hour FE                    & X                   & X                   & X           & X          \\
Day of Week FE             & X                    &                    & X            &           \\
Date FE                    &                     & X                   &             & X          \\
Location FE                & X                   & X                   & X           & X          \\
Observations               & 296,750             & 296,750             & 137,472     & 137,472    \\
R-squared                  & 0.403               & 0.035               & 0.558       & 0.000  \\\hline

\end{tabular}
\end{table}

The table reports the results of the trip-level regression of measures of detour distances on the number such trips a particular driver makes in 2013. The dependent variable is detour distance of LaGuardia or JFK to Manhattan trips. The independent variable is the number of LaGuardia or JFK to Manhattan trips that the driver of the given trip has made. Controls are included for the time and destination of the trips. Each observation is a single taxi ride. Standard errors are clustered at the driver level. *** p\textless{}0.01, ** p\textless{}0.05, * p\textless{}0.1

% Cultural Regressions: Severity
\newpage
\begin{table}[H]
\caption{Cultural Attitudes and Detour Distances}
\label{cultural_regs_severity}
\begin{tabular}{lcccc}
              & \multicolumn{2}{c}{LaGuardia} & \multicolumn{2}{c}{JFK} \\
                       & (1)           & (2)           & (3)        & (4)        \\\hline
CPI Score 2013         & 0.0420***     & 0.0395**     & 0.00503    & 0.00325    \\
                       & (0.0152)     & (0.0148)     & (0.00569)  & (0.00610)  \\
Constant               & 10.03***      & 7.792***      & 17.61***   & 10.24***   \\
                       & (0.0246)      & (0.0513)      & (0.0252)   & (0.212)    \\
                       &               &               &            &            \\
Controls (interacted): &               &               &            &            \\
Time FEs               & Yes           & No            & Yes        & No         \\
Destination FEs        & Yes           & No            & Yes        & No         \\
Google Maps Route      & No            & Yes           & No         & Yes        \\
                       &               &               &            &            \\
Observations           & 257,775       & 257,775       & 117,373    & 117,373    \\
R-squared              & 0.397         & 0.414         & 0.592      & 0.613      \\
Adj. R-squared         & 0.170         & 0.194         & 0.169      & 0.213      \\
Root MSE               & 1.171         & 1.154         & 1.211      & 1.179      \\\hline
                       &               &               &            &            \\
                       &               &               &            &           
\end{tabular}
\end{table}

This table presents the results of the trip-level regressions of the trip distances and Transparency International's Corruption Perceptions Index (CPI) scores in 2013 of the predicted home countries of the driver of the trip. CPI Score is computed following the methodology outlined in \hyperref[appendix-culture-data]{Appendix A}. The dependent variable is the distance of the trip. Time and destination controls are interacted. Google Maps Route is the distance of the counterfactual Google Maps recommended trip. Standard errors are clustered at the name-origin group level. *** p<0.01, ** p<0.05, * p<0.1.

% Cultural Regressions: Frequency
\newpage
\begin{table}[H]
\caption{Cultural Attitudes and Frequency of Trips}
\label{cultural_regs_frequency}
\begin{tabular}{lcc}
VARIABLES      & Number of LaGuardia Trips & Number of JFK Trips \\
               & (1)                       & (2)                 \\\hline
CPI Score 2013 & 0.295***                  & -0.430***           \\
               & (0.0651)                  & (0.0748)            \\
Constant       & 12.27***                  & 5.233***            \\
               & (0.246)                   & (0.257)             \\
               &                           &                     \\
Observations   & 22,824                    & 17,442              \\
R-squared      & 0.001                     & 0.003               \\
Adj. R-squared & 0.001                     & 0.002               \\
Root MSE       & 14.09                     & 10.81               \\\hline
               &                           &                     \\
               &                           &                    
\end{tabular}
\end{table}

This table presents the results of the driver-level regressions of the number of trips from a particular airport made by a driver and Transparency International's Corruption Perceptions Index (CPI) scores in 2013 of the predicted home countries of the driver. CPI Score is computed following the methodology outlined in \hyperref[appendix-culture-data]{Appendix A}. Robust standard errors are reported in parentheses. *** p<0.01, ** p<0.05, * p<0.1.

% Table 14A: Learning Regressions
\newpage
\begin{table}[H]
\caption{Learning of Detour Behavior over Time: All Drivers}
\label{learning_regs_fullsample}
\begin{tabular}{lccc}

                                                          & Trip Distance & In-Sample & Google Maps \\ \hline
                                                          & (1)           & (2)       & (3)         \\
                                                          &               &           &             \\
Driver Trip Number                                        & -0.0002       & -0.0003   & -0.001*     \\
                                                          & (0.0003)      & (0.0003)  & (0.0003)    \\
                                                          &               &           &             \\
Drivers FEs                                               & Yes           & Yes       & Yes         \\
                                                          &               &           &             \\
Observations                                              & 297,446       & 297,446   & 297,446     \\
R-squared                                                 & 0.437         & 0.461     & 0.315       \\
Adjusted R-squared                                        & 0.381         & 0.407     & 0.246       \\
Residual Std. Error                                       & 1.013         & 0.943     & 1.274       \\ \hline
*p\textless{}0.1; **p\textless{}0.05; ***p\textless{}0.01 &               &           &            
\end{tabular}
\end{table}

This table reports the results of the trip-level regression of detour distances on LaGuardia-to-Manhattan trips on the total number of previous such rides the driver of the given trip has taken since January 1, 2013. The dependent variable in column (1) is the trip distance. The dependent variable in column (2) is the trip distance residual. The dependent variable in column (3) is the Google Maps-based detour distance. Standard errors are clustered at the driver level.

% Table 14B: Learning Regressions
\newpage
\begin{landscape}
\begin{table}[H]
\caption{Learning of Detour Behavior over Time: New Drivers}
\label{learning_regs}
\begin{tabular}{lcccccc}
                                                          & \multicolumn{2}{c}{Post-March 1}            & \multicolumn{2}{c}{Post-June 1}             & \multicolumn{2}{c}{Post-Sep 1}              \\\hline
                                                          & \multicolumn{2}{c}{Detour Distance}         & \multicolumn{2}{c}{Detour Distance}         & \multicolumn{2}{c}{Detour Distance}         \\
                                                          & In-Sample            & Google Maps          & In-Sample            & Google Maps          & In-Sample            & Google Maps          \\
                                                          &                      &                      &                      &                      &                      &                      \\
Driver Trip Number                                        & 0.004**              & 0.003                & 0.007                & 0.007                & -0.011               & 0.008                \\
                                                          & (0.002)              & (0.002)              & (0.006)              & (0.006)              & (0.024)              & (0.025)              \\
                                                          &                      &                      &                      &                      &                      &                      \\
                                                          &                      &                      &                      &                      &                      &                      \\
Driver FEs                                                & Yes                  & Yes                  & Yes                  & Yes         & Yes                  & Yes                  \\
                                                          &                      &                      &                      &                      &                      &                      \\
Observations                                              & 16,986               & 18,344               & 5,070                & 5,504                & 1,130                & 1,237                \\
R-squared                                                 & 0.413                & 0.362                & 0.486                & 0.42                 & 0.568                & 0.48                 \\
Adjusted R-squared                                        & 0.286                & 0.231                & 0.317                & 0.238                & 0.324                & 0.203                \\
Residual Std. Error                                       & 1.170 (df = 13954)   & 1.318 (df = 15224)   & 1.145 (df = 3811)    & 1.309 (df = 4189)    & 1.122 (df = 722)     & 1.356 (df = 806)     \\\hline
\multicolumn{2}{l}{*p\textless{}0.1; **p\textless{}0.05; ***p\textless{}0.01}  & \multicolumn{1}{l}{} & \multicolumn{1}{l}{} & \multicolumn{1}{l}{} & \multicolumn{1}{l}{} & \multicolumn{1}{l}{}
\end{tabular}
\end{table}

This table reports the results of the trip-level regression of detour distances on LaGuardia-to-Manhattan trips on the total number of all previous taxi rides the driver of the given trip has taken since January 1, 2013. Post-March 1 is the subset of drivers who did not make a taxi trip in 2013 until March 1, 2013, which I consider plausibly new drivers. Post-June 1 and Post-Sep 1 are subsets defined analogously. Standard errors are clustered at the driver level.
\end{landscape}

%%% Figures

% Figure 1: Map

This figure provides a visualization of the areas of midtown Manhattan (yellow), LaGuardia airport (red), Residential area (grey), and JFK airport (green). Trips from (to) each of the areas are defined as trips that originate (end) in the respective colored rectangles. In the cases of the airports, the areas cover all the main taxi pick-up and drop-off lanes in the terminals.

% Figure 2: LGA Detour Histogram (Percentages)
\newpage

This figure plots a histogram of the detour distances of LaGuardia-to-midtown Manhattan trips. The horizontal axis is the detour distance as a percentage of the optimal Google Maps trip distance. The Google Maps distance is computed for a counterfactual trip with the same pick-up and drop-off latitudes and longitudes and the exact time of day and day of week of the trip. 

% Figure 3: JFK Detour Histogram (Percentages)
\newpage

This figure plots a histogram of the detour distances of JFK-to-midtown Manhattan trips. The horizontal axis is the detour distance as a percentage of the optimal Google Maps trip distance. The Google Maps distance is computed for a counterfactual trip with the same pick-up and drop-off latitudes and longitudes and the exact time of day and day of week of the trip. 

% Figures 4-5: Persistence Plots (In-Sample)
\newpage

The figure plots the correlation of average detour distances of LaGuardia to Manhattan taxi trips of each driver between the first six months and last six months of 2013. Each observation corresponds to a unique driver, whose average detour distance calculated using trips in first six months of 2013 corresponds to the value on the horizontal axis, and whose average detour distance calculated using trips in the last six months of 2013 corresponds to values on the vertical axis. Drivers are included in the sample if they have made at least 5 such trips in each of the two six-month periods of 2013. The average detour distances reported are the in-sample detour distance measure.

\newpage

The figure plots the correlation of average detour distances of LaGuardia to Manhattan taxi trips of each driver in the first six months of 2013 to the average detour distances of JFK to Manhattan taxi trips of the same driver in the last six months of 2013. Each observation corresponds to a unique driver, whose average detour distance calculated using LaGuardia trips in first six months of 2013 corresponds to the value on the horizontal axis, and whose average detour distance calculated using JFK trips in the last six months of 2013 corresponds to values on the vertical axis. Drivers are included in the sample if they have made at least 5 such trips in each of the two six-month periods of 2013. The average detour distances reported are the in-sample detour distance measure.

\newpage

The figure plots the correlation of average detour distances of LaGuardia to Manhattan taxi trips of each driver between the first six months and last six months of 2013. Each observation corresponds to a unique driver, whose average detour distance calculated using trips in first six months of 2013 corresponds to the value on the horizontal axis, and whose average detour distance calculated using trips in the last six months of 2013 corresponds to values on the vertical axis. Drivers are included in the sample if they have made at least 5 such trips in each of the two six-month periods of 2013. The average detour distances reported are the Google Maps-based detour distance measure.

% Figures 6-7: JFK Persistence Plots
\newpage

The figure plots the correlation of average detour distances of LaGuardia to Manhattan taxi trips of each driver in the first six months of 2013 to the average detour distances of JFK to Manhattan taxi trips of the same driver in the last six months of 2013. Each observation corresponds to a unique driver, whose average detour distance calculated using LaGuardia trips in first six months of 2013 corresponds to the value on the horizontal axis, and whose average detour distance calculated using JFK trips in the last six months of 2013 corresponds to values on the vertical axis. Drivers are included in the sample if they have made at least 5 such trips in each of the two six-month periods of 2013. The average detour distances reported are the Google Maps-based detour distance measure.

% Figures 8-9: Frequency Plots
\newpage

This plot is a binned plot of average trip distance binned by the number of trips a driver has made. The horizontal axis represents the number of LaGuardia to Manhattan trips driven by the driver in 2013. The vertical axis represents the average trip distance of LaGuardia to Manhattan trips of each driver. The plot is binned by the horizontal axis. The trip distance is measured in miles. The line of best fit is plotted weighted by the number of drivers.

This plot is a binned plot of average trip distance binned by the number of trips a driver has made. The horizontal axis represents the number of JFK to Manhattan trips driven by the driver in 2013. The vertical axis represents the average trip distance of JFK to Manhattan trips of each driver. The plot is binned by the horizontal axis. The trip distance is measured in miles. The line of best fit is plotted weighted by the number of drivers.

% Figures 10-11: Cultural Regressions Plots
\newpage

This figure presents the correlation between the Corruption Perceptions Index and the average detour distances of drivers on LaGuardia to Manhattan taxi trips at the name-origin group level for name-origin groups with at least 100 drivers. The Corruption Perceptions Index is weighted by the number of taxi drivers from each country in the name-origin group. Each observation represents a name-origin group and the point sizes are proportional to the number of drivers from each name-origin group. The line of best fit is plotted weighted by the number of drivers in each name-origin group.

This figure presents the correlation between the Corruption Perceptions Index and the average detour distances of drivers on JFK to Manhattan taxi trips at the name-origin group level for name-origin groups with at least 100 drivers. The Corruption Perceptions Index is weighted by the number of taxi drivers from each country in the name-origin group. Each observation represents a name-origin group and the point sizes are proportional to the number of drivers from each name-origin group. The line of best fit is plotted weighted by the number of drivers in each name-origin group.

%%%%% Appendix
\newpage
\setcounter{table}{0}
\setcounter{figure}{0}
\renewcommand{\thetable}{A\arabic{table}}
\renewcommand{\thefigure}{A\arabic{figure}}
\section*{Appendix}
\begin{appendices}
\newpage

% Cultural Norms Dataset Construction
\subsection*{Appendix A: Cultural Attitudes Dataset Construction}
\label{appendix_culture_data}
\subsubsection*{Name-Origin Groups}
\begin{figure}[H]
    \centering
    \caption{NamePrism Classification Taxonomy (from NamePrism)}
    \includegraphics[width=\textwidth]{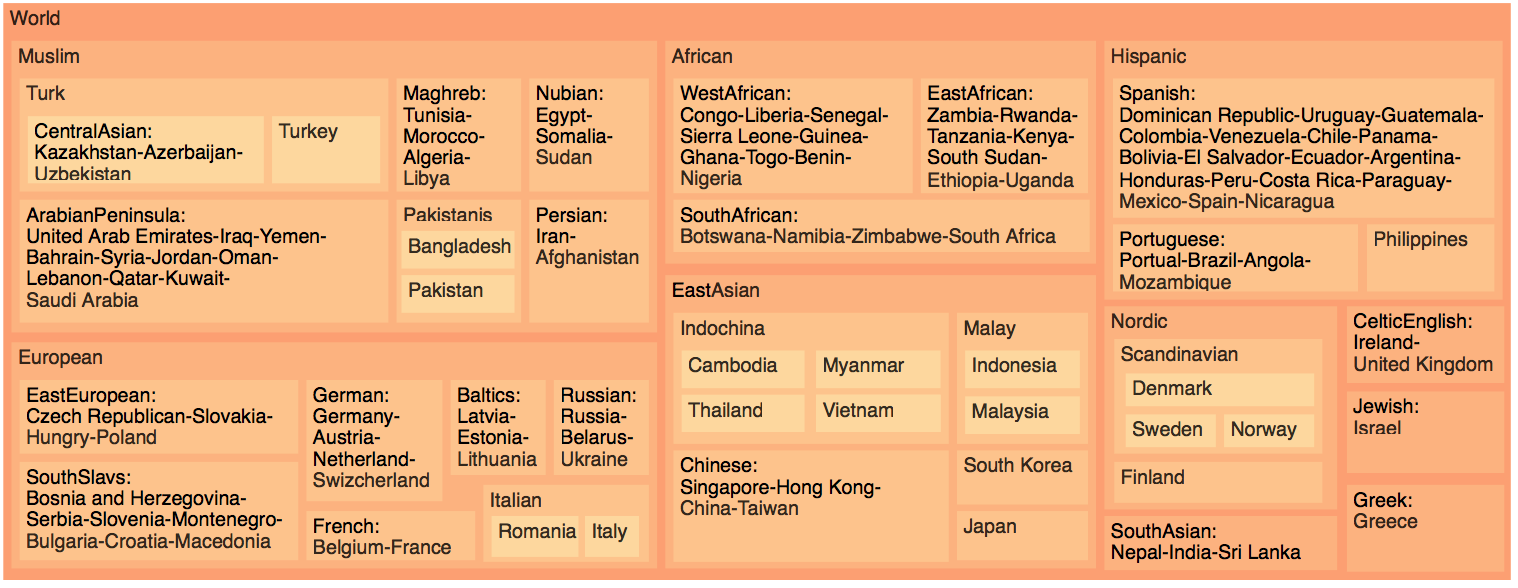}
    \label{nameprism_tree}
\end{figure}

I use the academic name-classification software NamePrism to match individual names to likely name-origin groups. The software computes the most probable name-origin group of an individual's name based on an algorithm trained on 74 million names from 118 countries. \autoref{nameprism_tree} provides a treemap of the classification outputs. I follow the NamePrism classification taxonomy and add Haiti and West African countries to the French label due to many drivers from these countries also having French names. I also add the US, which is not included as a name-origin group, as CelticEnglish. I perform robustness checks using a variety of weighting and inclusion methods for these countries.

\subsubsection*{Corruption Perceptions Index Score}
I use the Corruption Perceptions Index score from Transparency International to measure home country norms. I use 2013 scores in the main analysis and report robustness checks in \autoref{cultural_regs_severity_1998} and \autoref{cultural_regs_frequency_1998} using 1998 scores. I match countries to name-origin groups according to the procedure outlined in the Name-Origin Groups section above. I then match countries to administrative data from the TLC Factbook on the number of licensed taxi drivers who were born in each country. Countries with fewer than 10 drivers are removed because they are not reported in the TLC Factbook. I also remove countries that do not have 2013 CPI scores. The matched data is reported in \autoref{demographics-countries-scores}. 

% Table 5: Demographics Summary Stats
\newpage
\begin{table}[H]
\caption{Number of Drivers in Each Name-Origin Group}
\label{demographics-summary-stats}
\begin{tabular}{ll}
Name-Origin Group                         & Number of Drivers \\\hline
African,EastAfrican                       & 141               \\
African,SouthAfrican                      & 2                 \\
African,WestAfrican                       & 901               \\
CelticEnglish                             & 1814              \\
EastAsian,Chinese                         & 333               \\
EastAsian,Indochina,Myanmar               & 24                \\
EastAsian,Indochina,Thailand              & 12                \\
EastAsian,Indochina,Vietnam               & 99                \\
EastAsian,Japan                           & 19                \\
EastAsian,Malay,Indonesia                 & 441               \\
EastAsian,Malay,Malaysia                  & 72                \\
EastAsian,South Korea                     & 69                \\
European,Baltics                          & 2                 \\
European,EastEuropean                     & 48                \\
European,German                           & 175               \\
European,Italian,Italy                    & 41                \\
European,Italian,Romania                  & 129               \\
European,Russian                          & 341               \\
European,SouthSlavs                       & 46                \\
French                                    & 4548              \\
Greek                                     & 108               \\
Hispanic,Philippines                      & 105               \\
Hispanic,Portuguese                       & 141               \\
Hispanic,Spanish                          & 1010              \\
Jewish                                    & 24                \\
Muslim,ArabianPeninsula                   & 72                \\
Muslim,Maghreb                            & 42                \\
Muslim,Nubian                             & 1736              \\
Muslim,Pakistanis,Bangladesh              & 3944              \\
Muslim,Pakistanis,Pakistan                & 2693              \\
Muslim,Persian                            & 623               \\
Muslim,Turkic,CentralAsian                & 39                \\
Muslim,Turkic,Turkey                      & 155               \\
Nordic,Finland                            & 1                 \\
Nordic,Scandinavian,Norway                & 1                 \\
Nordic,Scandinavian,Sweden                & 4                 \\
SouthAsian                                & 4551              \\\hline
Total                                     & 24506            
\end{tabular}
\end{table}

\vspace{-2mm}

This table reports the number of drivers classified into each name-origin group by NamePrism. Only drivers who have made at least 1 LaGuardia or JFK trip in 2013 were included in the sample. The name-origin group taxonomy can be found in \hyperref[appendix-culture-data]{Appendix A}.

% Table 6: Demographics CPI Scores
\newpage
\begin{table}[H]
\caption{CPI Scores by Country}
\label{demographics-countries-scores}
\begin{tabular}{lllll}
Country              & Name-Origin Group            & \multicolumn{2}{c}{CPI Score}                       & Num. Drivers         \\
\multicolumn{1}{l}{} & \multicolumn{1}{l}{}         & \multicolumn{1}{l}{1998} & \multicolumn{1}{l}{2013} & \multicolumn{1}{l}{} \\\hline
Afghanistan          & Muslim,Persian               & -                        & 8                        & 139                  \\
Algeria              & Muslim,Maghreb               & -                        & 36                       & 387                  \\
Argentina            & Hispanic,Spanish             & 3                        & 34                       & 25                   \\
Azerbaijan           & Muslim,Turkic,CentralAsian   & -                        & 28                       & 24                   \\
Bangladesh           & Muslim,Pakistanis,Bangladesh & -                        & 27                       & 7684                 \\
Belarus              & European,Russian             & 3.9                      & 29                       & 22                   \\
Benin                & African,WestAfrican          & -                        & 36                       & 26                   \\
Bolivia              & Hispanic,Spanish             & 2.8                      & 34                       & 18                   \\
Brazil               & Hispanic,Portuguese          & 4                        & 42                       & 36                   \\
Bulgaria             & European,SouthSlavs          & 2.9                      & 41                       & 74                   \\
China                & EastAsian,Chinese            & 3.5                      & 40                       & 441                  \\
Colombia             & Hispanic,Spanish             & 2.2                      & 36                       & 212                  \\
Congo                & African,WestAfrican          & -                        & 22                       & 23                   \\
Dominican Republic   & Hispanic,Spanish             & -                        & 29                       & 349                  \\
Ecuador              & Hispanic,Spanish             & 2.3                      & 35                       & 288                  \\
Egypt                & Muslim,Nubian                & 2.9                      & 32                       & 1076                 \\
El Salvador          & Hispanic,Spanish             & 3.6                      & 38                       & 26                   \\
Ethiopia             & African,EastAfrican          & -                        & 33                       & 104                  \\
France               & French                       & 6.7                      & 71                       & 15                   \\
Germany              & European,German              & 7.9                      & 78                       & 10                   \\
Ghana                & French                       & 3.3                      & 46                       & 1300                 \\
Greece               & Greek                        & 4.9                      & 40                       & 133                  \\
Guatemala            & Hispanic,Spanish             & 3.1                      & 29                       & 15                   \\
Guinea               & French                       & -                        & 24                       & 494                  \\
Haiti                & French                       & -                        & 19                       & 1857                 \\
Honduras             & Hispanic,Spanish             & 1.7                      & 26                       & 15                   \\
Hungary              & European,EastEuropean        & 5                        & 54                       & 19                   \\
India                & SouthAsian                   & 2.9                      & 36                       & 2593                 \\
Indonesia            & EastAsian,Malay,Indonesia    & 2                        & 32                       & 84                   \\
Iran                 & Muslim,Persian               & -                        & 25                       & 35                   \\
Israel               & Jewish                       & 7.1                      & 61                       & 35                   \\
Italy                & European,Italian,Italy       & 4.6                      & 43                       & 17                   \\
Japan                & EastAsian,Japan              & 5.8                      & 74                       & 10                   \\
Jordan               & Muslim,ArabianPeninsula      & 4.7                      & 45                       & 42                   \\
Kazakhstan           & Muslim,Turkic,CentralAsian   & -                        & 26                       & 14                   \\
Kuwait               & Muslim,ArabianPeninsula      & -                        & 43                       & 10                   \\
Lebanon              & Muslim,ArabianPeninsula      & -                        & 28                       & 45                   \\
Liberia              & French                       & -                        & 38                       & 82                   \\
Malaysia             & EastAsian,Malay,Malaysia     & 5.3                      & 50                       & 14                   \\
\end{tabular}
\end{table}

\pagebreak
\newpage

\begin{tabular}{lllll}
Country              & Name-Origin Group            & \multicolumn{2}{c}{CPI Score}                       & Num. Drivers         \\
\multicolumn{1}{l}{} & \multicolumn{1}{l}{}         & \multicolumn{1}{l}{1998} & \multicolumn{1}{l}{2013} & \multicolumn{1}{l}{} \\\hline
Mexico               & Hispanic,Spanish             & 3.3                      & 34                       & 34                   \\
Morocco              & Muslim,Maghreb               & 3.7                      & 37                       & 817                  \\
Myanmar              & EastAsian,Indochina,Myanmar  & -                        & 21                       & 85                   \\
Nepal                & SouthAsian                   & -                        & 31                       & 472                  \\
Nigeria              & French                       & 1.9                      & 25                       & 412                  \\
Pakistan             & Muslim,Pakistanis,Pakistan   & 2.7                      & 28                       & 3125                 \\
Panama               & Hispanic,Spanish             & -                        & 35                       & 12                   \\
Peru                 & Hispanic,Spanish             & 4.5                      & 38                       & 88                   \\
Philippines          & Hispanic,Philippines         & 3.3                      & 36                       & 42                   \\
Poland               & European,EastEuropean        & 4.6                      & 60                       & 95                   \\
Romania              & European,Italian,Romania     & 3                        & 43                       & 154                  \\
Russia               & European,Russian             & 2.4                      & 28                       & 203                  \\
Senegal              & French                       & 3.3                      & 41                       & 447                  \\
Sierra Leone         & French                       & -                        & 30                       & 114                  \\
Somalia              & Muslim,Nubian                & -                        & 8                        & 22                   \\
South Korea          & EastAsian,South Korea        & 4.2                      & 55                       & 107                  \\
Sri Lanka            & SouthAsian                   & -                        & 37                       & 51                   \\
Sudan                & Muslim,Nubian                & -                        & 11                       & 218                  \\
Syria                & Muslim,ArabianPeninsula      & -                        & 17                       & 25                   \\
Taiwan               & EastAsian,Chinese            & 5.3                      & 61                       & 17                   \\
Thailand             & EastAsian,Indochina,Thailand & 3                        & 35                       & 19                   \\
Togo                 & African,WestAfrican          & -                        & 29                       & 89                   \\
Tunisia              & Muslim,Maghreb               & 5                        & 41                       & 60                   \\
Turkey               & Muslim,Turkic,Turkey         & 3.4                      & 50                       & 176                  \\
Ukraine              & European,Russian             & 2.8                      & 25                       & 85                   \\
United States        & CelticEnglish                & 7.5                      & 73                       & 1303                 \\
Uzbekistan           & Muslim,Turkic,CentralAsian   & -                        & 17                       & 189                  \\
Vietnam              & EastAsian,Indochina,Vietnam  & 2.5                      & 31                       & 80                   \\
Yemen                & Muslim,ArabianPeninsula      & -                        & 18                       & 55                  

\end{tabular}

\vspace{0.1in}

This table presents the Corruption Perceptions Index (CPI) scores of driver home countries by name-origin group and the number of drivers from each country. CPI scores are from Transparency International. Name-origin groups are based on the NamePrism classification taxonomy. The number of drivers is from NYC TLC administrative data.

% Cultural Norms: Standard Errors
\newpage
\begin{table}[H]
\caption{Cultural Attitudes and Detour Distances (Bootstrapped Standard Errors)}
\label{cultural_regs_severity_clusterse}
\begin{tabular}{lcccc}
VARIABLES              & \multicolumn{2}{c}{LaGuardia}    & \multicolumn{2}{c}{JFK}      \\ \hline
                       &                 &                &               &              \\
CPI Score 2013         & 0.042           & 0.0395         & 0.00503       & 0.00325      \\
                       & (0.00692)***    & (0.00674)***   & (0.00703)     & (0.00668)    \\
                       & {[}0.0179{]}*** & {[}0.0193{]}** & {[}0.00431{]} & {[}0.0121{]} \\
                       & \{0.0152\}***   & \{0.0148\}**  & \{0.00569\}   & \{0.00610\}  \\
                       &                 &                &               &              \\
Controls (interacted): &                 &                &               &              \\
Time FEs               & Yes             & No             & Yes           & No           \\
Destination FEs        & Yes             & No             & Yes           & No           \\
Google Maps Route      & No              & Yes            & No            & Yes          \\
                       &                 &                &               &              \\
Observations           & 257,775         & 257,775        & 117,373       & 117,373      \\
R-squared              & 0.397           & 0.414          & 0.592         & 0.613        \\
Adj. R-squared         & 0.170           & 0.194          & 0.169         & 0.213        \\
Root MSE               & 1.171           & 1.154          & 1.211         & 1.179       \\ \hline
\end{tabular}
\end{table}

This table presents the results of the trip-level regressions of the trip distances and Transparency International's Corruption Perceptions Index (CPI) scores in 2013 of the predicted home countries of the driver of the trip using various standard error computations. () report standard errors computed with clustering at the driver level. [] report standard errors computed using the wild cluster boostrap following \cite{cameron2008bootstrap}, correcting for robust number of clusters at the name-origin group level. \{\} report standard errors computed using clustering at the name-origin group level. CPI Score is computed following the methodology outlined in \hyperref[appendix-culture-data]{Appendix A}. The dependent variable is the distance of the trip. Time and destination controls are interacted. Google Maps Route is the distance of the counterfactual Google Maps recommended trip. *** p<0.01, ** p<0.05, * p<0.1.

% Cultural Norms: Remove-One Regs
\newpage
\begin{landscape}
\begin{table}[H]
\centering
\caption{Cultural Norms: Remove-One Regressions}
\label{ethnicity_removeone_regs}

% \usepackage{multirow}
\begin{tabular}{lcccccc} \hline
               & (1)         & (2)          & (3)         & (4)           & (5)       & (6)       \\
Removed:               & EastAfrican & SouthAfrican & WestAfrican & CelticEnglish & Chinese   & Myanmar   \\
               &             &              &             &               &           &           \\ 
CPI Score 2013 & 0.0420***   & 0.0419***    & 0.0409***   & 0.0534***     & 0.0409*** & 0.0429*** \\
               & (0.00693)   & (0.00692)    & (0.00692)   & (0.0116)      & (0.00694) & (0.00693) \\
Constant       & 10.03***    & 10.03***     & 10.03***    & 10.06***      & 10.03***  & 10.04***  \\
               & (0.0246)    & (0.0246)     & (0.0245)    & (0.0366)      & (0.0245)  & (0.0246)  \\
               &             &              &             &               &           &           \\
Observations   & 256,560     & 257,768      & 250,221     & 241,721       & 254,383   & 257,409   \\
R-squared      & 0.398       & 0.397        & 0.404       & 0.410         & 0.399     & 0.398     \\
Adj. R-squared & 0.170       & 0.170        & 0.172       & 0.173         & 0.170     & 0.170     \\
Root MSE       & 1.171       & 1.171        & 1.168       & 1.167         & 1.171     & 1.171    \\ \hline

               & (7)       & (8)       & (9)       & (10)      & (11)      & (12)        \\
Removed:               & Thailand  & Vietnam   & Japan     & Indonesia & Malaysia  & South Korea \\
               &           &           &           &           &           &             \\
CPI Score 2013 & 0.0419*** & 0.0422*** & 0.0415*** & 0.0420*** & 0.0419*** & 0.0407***   \\
               & (0.00692) & (0.00693) & (0.00695) & (0.00691) & (0.00695) & (0.00694)   \\
Constant       & 10.03***  & 10.03***  & 10.03***  & 10.03***  & 10.03***  & 10.03***    \\
               & (0.0246)  & (0.0246)  & (0.0246)  & (0.0246)  & (0.0246)  & (0.0246)    \\
               &           &           &           &           &           &             \\
Observations   & 257,677   & 256,795   & 257,651   & 253,442   & 256,970   & 256,995     \\
R-squared      & 0.397     & 0.398     & 0.398     & 0.400     & 0.398     & 0.398       \\
Adj. R-squared & 0.170     & 0.170     & 0.170     & 0.170     & 0.170     & 0.171       \\
Root MSE       & 1.171     & 1.171     & 1.171     & 1.171     & 1.171     & 1.171    \\ \hline  
\end{tabular}
\end{table}

\newpage
\begin{table}[H]
\begin{tabular}{lcccccc}\hline
               & (13)      & (14)         & (15)      & (16)      & (17)      & (18)         \\
Removed:       & Baltics   & EastEuropean & French    & German    & Italy     & SouthAfrican \\
               &           &              &           &           &           &              \\
CPI Score 2013 & 0.0419*** & 0.0417***    & 0.0523*** & 0.0446*** & 0.0418*** & 0.0419***    \\
               & (0.00692) & (0.00696)    & (0.00697) & (0.00728) & (0.00693) & (0.00692)    \\
Constant       & 10.03***  & 10.03***     & 10.12***  & 10.04***  & 10.03***  & 10.03***     \\
               & (0.0246)  & (0.0246)     & (0.0252)  & (0.0255)  & (0.0246)  & (0.0246)     \\
               &           &              &           &           &           &              \\
Observations   & 257,770   & 257,349      & 225,826   & 256,451   & 257,491   & 257,768      \\
R-squared      & 0.397     & 0.398        & 0.422     & 0.399     & 0.398     & 0.397        \\
Adj. R-squared & 0.170     & 0.170        & 0.175     & 0.171     & 0.170     & 0.170        \\
Root MSE       & 1.171     & 1.171        & 1.161     & 1.170     & 1.171     & 1.171      \\ \hline

               & (19)      & (20)       & (21)      & (22)        & (23)       & (24)      \\
Removed:       & Russian   & SouthSlavs & Greek     & Philippines & Portuguese & Spanish   \\
               &           &            &           &             &            &           \\
CPI Score 2013 & 0.0407*** & 0.0418***  & 0.0426*** & 0.0418***   & 0.0417***  & 0.0420*** \\
               & (0.00693) & (0.00692)  & (0.00691) & (0.00692)   & (0.00693)  & (0.00694) \\
Constant       & 10.02***  & 10.03***   & 10.03***  & 10.03***    & 10.03***   & 10.03***  \\
               & (0.0246)  & (0.0246)   & (0.0246)  & (0.0246)    & (0.0246)   & (0.0245)  \\
               &           &            &           &             &            &           \\
Observations   & 255,040   & 257,446    & 255,573   & 256,881     & 256,412    & 244,300   \\
R-squared      & 0.400     & 0.398      & 0.399     & 0.398       & 0.398      & 0.407     \\
Adj. R-squared & 0.172     & 0.170      & 0.170     & 0.170       & 0.170      & 0.172     \\
Root MSE       & 1.169     & 1.171      & 1.169     & 1.170       & 1.171      & 1.165    \\ \hline
\end{tabular}
\end{table}

\newpage
\begin{table}[H]
\begin{tabular}{lccccccc}\hline
               & (25)      & (26)             & (27)      & (28)      & (29)       & (30)      &  \\
Removed:       & Jewish    & ArabianPeninsula & Maghreb   & Nubian    & Bangladesh & Pakistan  &  \\
               &           &                  &           &           &            &           &  \\
CPI Score 2013 & 0.0427*** & 0.0421***        & 0.0419*** & 0.0446*** & 0.0271***  & 0.0382*** &  \\
               & (0.00694) & (0.00693)        & (0.00692) & (0.00699) & (0.00732)  & (0.00705) &  \\
Constant       & 10.03***  & 10.03***         & 10.03***  & 10.05***  & 9.950***   & 10.01***  &  \\
               & (0.0246)  & (0.0246)         & (0.0246)  & (0.0251)  & (0.0277)   & (0.0254)  &  \\
               &           &                  &           &           &            &           &  \\
Observations   & 257,565   & 256,996          & 257,599   & 244,290   & 204,251    & 232,643   &  \\
R-squared      & 0.398     & 0.398            & 0.397     & 0.408     & 0.435      & 0.415     &  \\
Adj. R-squared & 0.170     & 0.170            & 0.170     & 0.174     & 0.170      & 0.172     &  \\
Root MSE       & 1.171     & 1.171            & 1.171     & 1.168     & 1.182      & 1.172     & \\ \hline

               & (31)      & (32)         & (33)      & (34)      & (35)      & (36)      & (37)       \\
Removed:       & Persian   & CentralAsian & Turkey    & Finland   & Norway    & Sweden    & SouthAsian \\
               &           &              &           &           &           &           &            \\
CPI Score 2013 & 0.0405*** & 0.0419***    & 0.0412*** & 0.0420*** & 0.0420*** & 0.0418*** & 0.0448***  \\
               & (0.00745) & (0.00693)    & (0.00695) & (0.00693) & (0.00692) & (0.00693) & (0.00712)  \\
Constant       & 10.03***  & 10.03***     & 10.03***  & 10.03***  & 10.03***  & 10.03***  & 10.01***   \\
               & (0.0270)  & (0.0246)     & (0.0246)  & (0.0246)  & (0.0246)  & (0.0246)  & (0.0252)   \\
               &           &              &           &           &           &           &            \\
Observations   & 250,180   & 257,469      & 256,816   & 257,770   & 257,774   & 257,754   & 193,838    \\
R-squared      & 0.402     & 0.398        & 0.398     & 0.397     & 0.397     & 0.397     & 0.435      \\
Adj. R-squared & 0.170     & 0.170        & 0.170     & 0.170     & 0.170     & 0.170     & 0.156      \\
Root MSE       & 1.173     & 1.171        & 1.171     & 1.171     & 1.171     & 1.171     & 1.183     \\ \hline
\end{tabular}
\end{table}

This table presents the results of the trip-level regressions of the trip distances and Transparency International's Corruption Perceptions Index (CPI) scores in 2013 of the predicted home countries of the driver of the trip, with the sample restricted to trips made by all drivers except those from predicted home countries in the name-origin group removed. Each regression follows the specification of column (1) in \autoref{cultural_regs_severity}. CPI Score is computed following the methodology outlined in \hyperref[appendix-culture-data]{Appendix A}. The dependent variable is the distance of the trip. Time and destination controls are interacted. Standard errors are clustered at the driver level. *** p<0.01, ** p<0.05, * p<0.1.

\end{landscape}

% Cultural Norms: 1998 CPI Score
\newpage
\begin{table}[H]
\caption{Cultural Attitudes and Detour Distances (CPI 1998)}
\label{cultural_regs_severity_1998}
\begin{tabular}{lcccc}
VARIABLES              & \multicolumn{2}{c}{LaGuardia} & \multicolumn{2}{c}{JFK} \\
                       & (1)           & (2)           & (3)        & (4)        \\\hline
CPI Score 1998         & 0.0281***     & 0.0277***     & -0.00156   & -0.00329   \\
                       & (0.00756)     & (0.00739)     & (0.00704)  & (0.00666)  \\
Constant               & 9.944***      & 7.648***      & 17.57***   & 10.05***   \\
                       & (0.0274)      & (0.0605)      & (0.0246)   & (0.243)    \\
                       &               &               &            &            \\
Controls (interacted): &               &               &            &            \\
Time FEs               & Yes           & No            & Yes        & No         \\
Destination FEs        & Yes           & No            & Yes        & No         \\
Google Maps Route      & No            & Yes           & No         & Yes        \\
                       &               &               &            &            \\
Observations           & 195,984       & 195,984       & 102,194    & 102,194    \\
R-squared              & 0.442         & 0.458         & 0.622      & 0.642      \\
Adj. R-squared         & 0.170         & 0.194         & 0.172      & 0.217      \\
Root MSE               & 1.185         & 1.167         & 1.206      & 1.173     \\\hline
\end{tabular}
\end{table}

This table presents the results of the trip-level regressions of the trip distances and Transparency International's Corruption Perceptions Index (CPI) scores in 1998 of the predicted home countries of the driver of the trip. CPI Score is computed following the methodology outlined in \hyperref[appendix-culture-data]{Appendix A}. The dependent variable is the distance of the trip. Time and destination controls are interacted. Google Maps Route is the distance of the counterfactual Google Maps recommended trip. Standard errors are clustered at the driver level. *** p<0.01, ** p<0.05, * p<0.1.

\newpage
\begin{table}[H]
\caption{Cultural Attitudes and Frequency of Trips (CPI 1998)}
\label{cultural_regs_frequency_1998}
\begin{tabular}{lcc}
VARIABLES      & Number of LaGuardia Trips & Number of JFK Trips \\
               & (1)                       & (2)                 \\ \hline
CPI Score 1998 & 0.248***                  & -0.111              \\
               & (0.0649)                  & (0.0769)            \\
Constant       & 11.49***                  & 6.771***            \\
               & (0.251)                   & (0.269)             \\
               &                           &                     \\
Observations   & 18,371                    & 14,269              \\
R-squared      & 0.001                     & 0.000               \\
Adj. R-squared & 0.000                     & 0.000               \\
Root MSE       & 13.77                     & 11.44               \\ \hline
\end{tabular}
\end{table}

This table presents the results of the driver-level regressions of the number of trips from a particular airport made by a driver and Transparency International's Corruption Perceptions Index (CPI) scores in 1998 of the predicted home countries of the driver. CPI Score is computed following the methodology outlined in \hyperref[appendix-culture-data]{Appendix A}. Robust standard errors are reported in parentheses. *** p<0.01, ** p<0.05, * p<0.1.

% Cultural Norms: Unweighted
\newpage
\begin{table}[H]
\caption{Cultural Attitudes and Detour Distances (CPI Unweighted)}
\label{cultural_regs_severity_unweighted}
\begin{tabular}{lcccc}
                       & LaGuardia &           & JFK       &           \\
                       & (1)       & (2)       & (3)       & (4)       \\ \hline
CPI Score 2013         & 0.0505*** & 0.0478*** & 0.000663  & -0.000636 \\
                       & (0.00645) & (0.00628) & (0.00670) & (0.00636) \\
Constant               & 10.06***  & 7.828***  & 17.59***  & 10.23***  \\
                       & (0.0233)  & (0.0506)  & (0.0246)  & (0.212)   \\
                       &           &           &           &           \\
Controls (interacted): &           &           &           &           \\
Time FEs               & Yes       & No        & Yes       & No        \\
Destination FEs        & Yes       & No        & Yes       & No        \\
Google Maps Route      & No        & Yes       & No        & Yes       \\
                       &           &           &           &           \\
Observations           & 257,775   & 257,775   & 117,373   & 117,373   \\
R-squared              & 0.398     & 0.415     & 0.592     & 0.613     \\
Adj. R-squared         & 0.171     & 0.194     & 0.169     & 0.213     \\
Root MSE               & 1.170     & 1.154     & 1.211     & 1.179     \\ \hline
\end{tabular}
\end{table}

This table presents the results of the trip-level regressions of the trip distances and Transparency International's Corruption Perceptions Index (CPI) scores in 2013 of the predicted home countries of the driver of the trip. CPI Score is computed following the methodology outlined in \hyperref[appendix-culture-data]{Appendix A} as an unweighted simple average. The dependent variable is the distance of the trip. Time and destination controls are interacted. Google Maps Route is the distance of the counterfactual Google Maps recommended trip. Standard errors are clustered at the driver level. *** p<0.01, ** p<0.05, * p<0.1.

\newpage
\begin{table}[H]
\caption{Cultural Attitudes and Frequency of Trips (CPI Unweighted)}
\label{cultural_regs_frequency_unweighted}

\begin{tabular}{lcc}
VARIABLES      & Number of LaGuardia Trips & Number of JFK Trips \\
               & (1)                       & (2)                 \\ \hline
CPI Score 2013 & 0.305***                  & -0.495***           \\
               & (0.0600)                  & (0.0658)            \\
Constant       & 12.33***                  & 4.965***            \\
               & (0.234)                   & (0.229)             \\
               &                           &                     \\
Observations   & 22,824                    & 17,442              \\
R-squared      & 0.001                     & 0.004               \\
Adj. R-squared & 0.000849                  & 0.00388             \\
Root MSE       & 14.09                     & 10.80               \\ \hline
\end{tabular}
\end{table}

This table presents the results of the driver-level regressions of the number of trips from a particular airport made by a driver and Transparency International's Corruption Perceptions Index (CPI) scores in 2013 of the predicted home countries of the driver. CPI Score is computed following the methodology outlined in \hyperref[appendix-culture-data]{Appendix A} as an unweighted simple average. Robust standard errors are reported in parentheses. *** p<0.01, ** p<0.05, * p<0.1.

\newpage
\begin{table}[H]
\caption{Cultural Attitudes and Detour Distances (Probability-Weighted Name Mapping)}
\label{cultural_regs_severity_proportional}
\begin{tabular}{lcccccc} \hline
 & (1) & (2) & (3) & (4) & (5) & (6) \\
VARIABLES & &  & &  & & \\ \hline
 &  &  &  &  &  &  \\
CPI Score 2013 & 0.054*** & 0.037**  &  0.064** & 0.077*** & 0.050**  & 0.15*** \\
 & (0.018) & (0.015)  & (0.028)  & (0.019)  & (0.023) & (0.040) \\
Constant & 10.1*** & 10.0*** & 10.2*** & 10.2*** & 10.1*** & 10.4*** \\
 & (0.076) & (0.082) & (0.094) & (0.092) & (0.12) & (0.12) \\
 &  &  &  &  &  &  \\
Observations & 234,297 & 234,297 & 234,297 & 234,297 & 234,297 & 234,297 \\
R-squared & 0.337 & 0.335 & 0.372 & 0.337 & 0.335 & 0.338 \\
Adj. R-squared & 0.16 & 0.16 & 0.17 & 0.16 & 0.16 & 0.17 \\
Root MSE & 1.17 & 1.17 & 1.14 & 1.17 & 1.17 & 1.17 \\ \hline
\end{tabular}
\end{table}

This table presents the results of the trip-level regressions of the trip distances and Transparency International's Corruption Perceptions Index (CPI) scores in 2013 of the predicted home countries of the driver of the trip using various standard error computations. CPI Score is computed following the methodology outlined in \hyperref[appendix-culture-data]{Appendix A}. Columns 1-3 use the original maximum probability (one-to-one) for the mapping, whereas columns 4-6 use the probability-weighted mapping. Columns 1 and 4 use the simple average across home countries; columns 2 and 5 use the median across home countries; columns 3 and 6 use the weighted average weighted by the number of drivers from each country. The dependent variable is the distance of the trip. Time and destination controls are interacted. Standard errors are clustered at the name-origin group level. *** p<0.01, ** p<0.05, * p<0.1.

\newpage
% latex table generated in R 3.6.1 by xtable 1.8-4 package
% Mon Jan 06 21:20:00 2020
\begin{table}[ht]
\caption{Proportion of Drivers by Airport}
\label{driver_counts_byairport}
\centering
\begin{tabular}{rlrr}
  \hline
 & Name-Origin Group & \% of LGA Drivers & \% of JFK Drivers \\ 
  \hline
1 & SouthAsian & 0.19 & 0.20 \\ 
  2 & French & 0.18 & 0.19 \\ 
  3 & Muslim,Pakistanis,Bangladesh & 0.16 & 0.15 \\ 
  4 & Muslim,Pakistanis,Pakistan & 0.11 & 0.11 \\ 
  5 & CelticEnglish & 0.07 & 0.07 \\ 
  6 & Muslim,Nubian & 0.07 & 0.06 \\ 
  7 & Hispanic,Spanish & 0.04 & 0.03 \\ 
  8 & African,WestAfrican & 0.04 & 0.03 \\ 
  9 & Muslim,Persian & 0.03 & 0.02 \\ 
  10 & EastAsian,Malay,Indonesia & 0.02 & 0.02 \\ 
  11 & Unknown & 0.02 & 0.02 \\ 
  12 & EastAsian,Chinese & 0.01 & 0.01 \\ 
  13 & European,Russian & 0.01 & 0.02 \\ 
  14 & European,German & 0.01 & 0.01 \\ 
  15 & Hispanic,Portuguese & 0.01 & 0.01 \\ 
  16 & Muslim,Turkic,Turkey & 0.01 & 0.01 \\ 
  17 & African,EastAfrican & 0.01 & 0.01 \\ 
  18 & European,Italian,Romania & 0.01 & 0.01 \\ 
  19 & Greek & 0.00 & 0.00 \\ 
  20 & EastAsian,Indochina,Vietnam & 0.00 & 0.00 \\ 
  21 & Hispanic,Philippines & 0.00 & 0.00 \\ 
  22 & EastAsian,Malay,Malaysia & 0.00 & 0.00 \\ 
  23 & Muslim,ArabianPeninsula & 0.00 & 0.00 \\ 
  24 & EastAsian,South Korea & 0.00 & 0.00 \\ 
  25 & European,EastEuropean & 0.00 & 0.00 \\ 
  26 & European,SouthSlavs & 0.00 & 0.00 \\ 
  27 & Muslim,Turkic,CentralAsian & 0.00 & 0.00 \\ 
  28 & European,Italian,Italy & 0.00 & 0.00 \\ 
  29 & Muslim,Maghreb & 0.00 & 0.00 \\ 
  30 & EastAsian,Indochina,Myanmar & 0.00 & 0.00 \\ 
  31 & Jewish & 0.00 & 0.00 \\ 
  32 & EastAsian,Japan & 0.00 & 0.00 \\ 
  33 & EastAsian,Indochina,Thailand & 0.00 & 0.00 \\ 
  34 & Nordic,Scandinavian,Sweden & 0.00 & 0.00 \\ 
  35 & African,SouthAfrican & 0.00 & 0.00 \\ 
  36 & European,Baltics & 0.00 & 0.00 \\ 
  37 & Nordic,Finland & 0.00 & 0.00 \\ 
  38 & Nordic,Scandinavian,Norway & 0.00 &  \\ 
   \hline
\end{tabular}
\end{table}

% Random Split Plots
\newpage
\begin{figure}[H]
    \centering
    \caption{Persistence of Detour Distances (Google Maps): Random Split}
    \includegraphics[height=260pt]{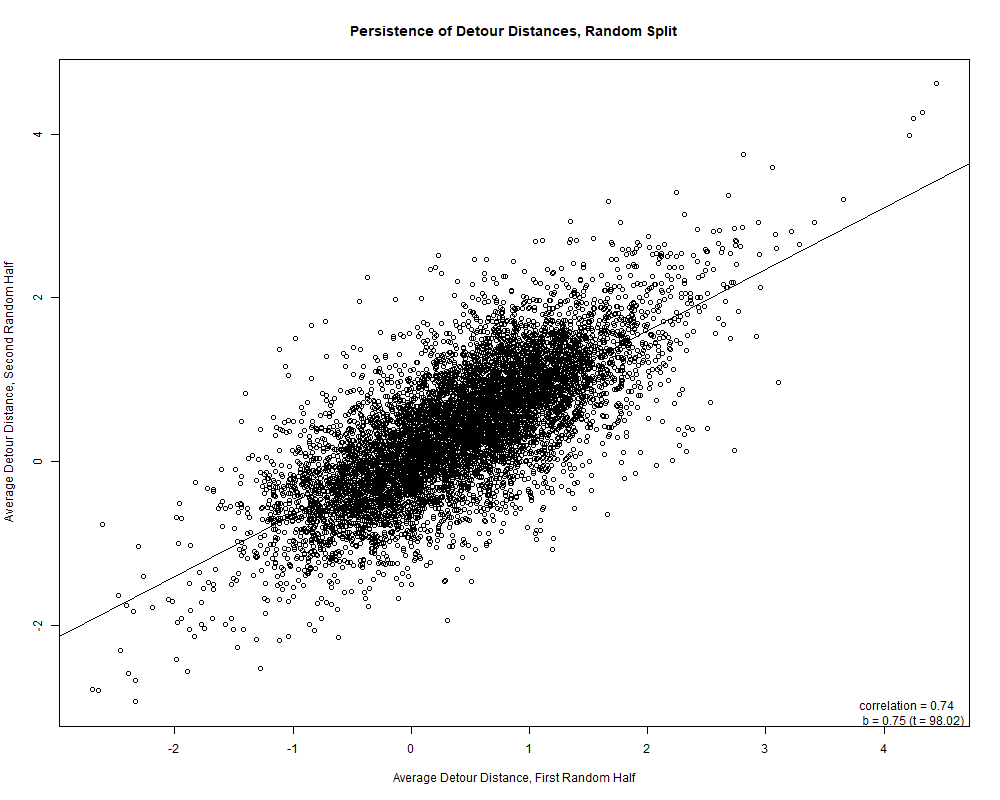}
    \label{persistence_gmap_LGALGA_rand_LGA1HLGA2H}
\end{figure}

\begin{figure}[H]
    \centering
    \caption{Persistence of Detour Distances (In-Sample): Random Split}
    \includegraphics[height=260pt]{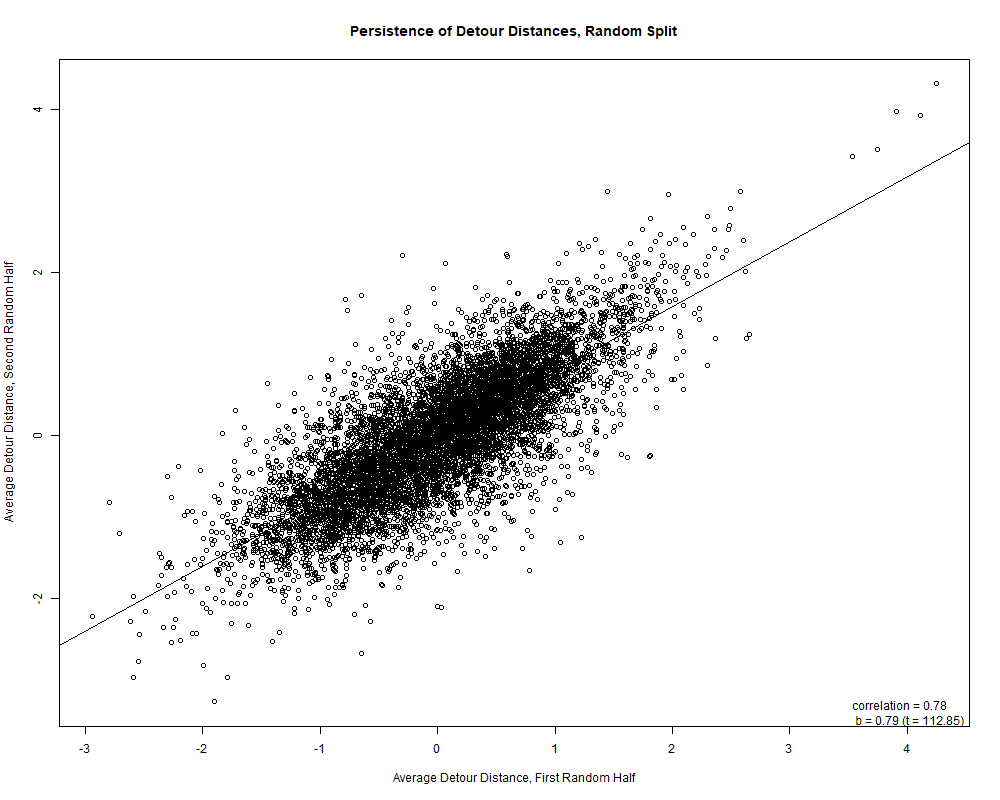}
    \label{persistence_insample_LGALGA_rand_LGA1HLGA2H}
\end{figure}

% Marginal Utility of Money
% Figure 12: Marginal Utility - Christmas Plot
\newpage
\begin{figure}[H]
    \centering
    \caption{Marginal Utility: Christmas}
    \includegraphics[height=280pt]{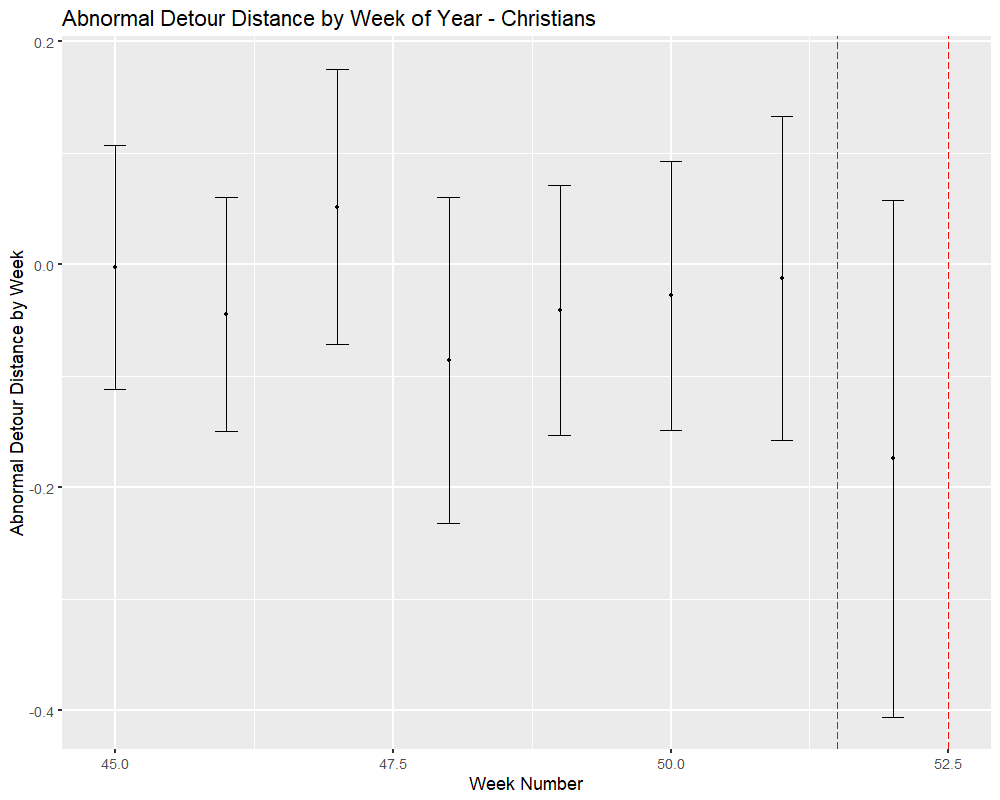}
    \label{marginalutility_christians}
\end{figure}

% Figure 13: Marginal Utility - CNY Plot
\newpage
\begin{figure}[H]
    \centering
    \caption{Marginal Utility: Lunar New Year}
    \includegraphics[height=280pt]{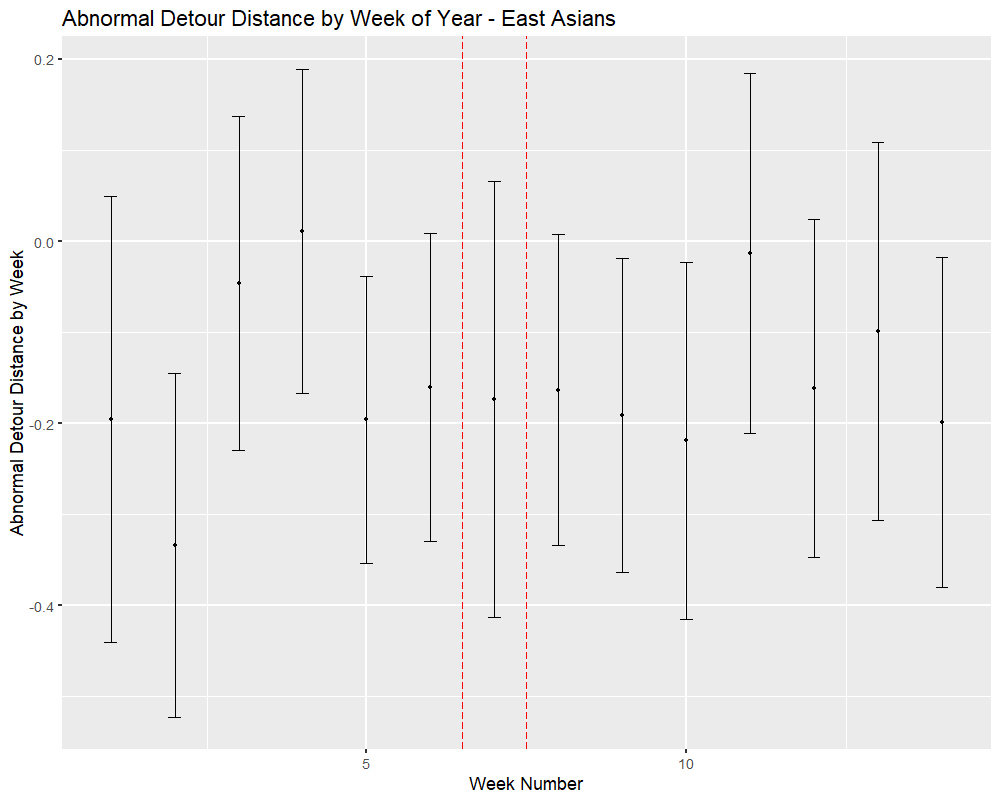}
    \label{marginalutility_cny}
\end{figure}

% Figure 14: Marginal Utility - Ramadan Plot
\newpage
\begin{figure}[H]
    \centering
    \caption{Marginal Utility: Ramadan}
    \includegraphics[height=280pt]{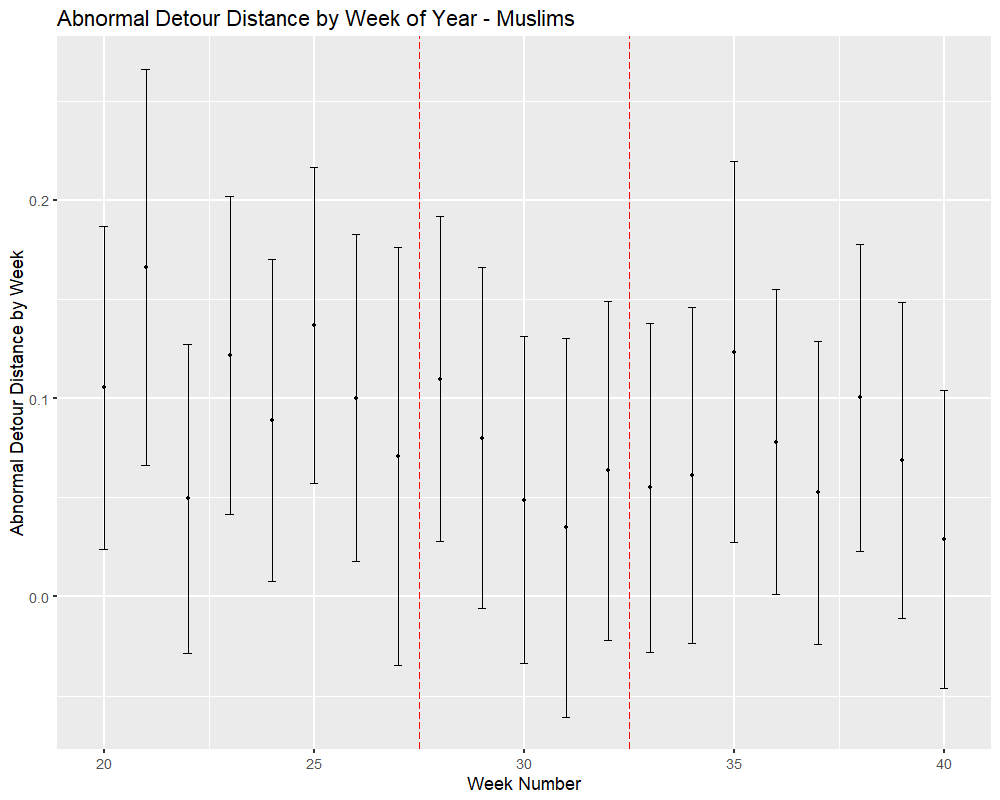}
    \label{marginalutility_muslims}
\end{figure}

% Figure 15: Learning Linear Curve: All Drivers
\newpage
\begin{figure}[H]
    \centering
    \caption{Learning of Detour Behaviors: All Drivers}
    \includegraphics[height=280pt]{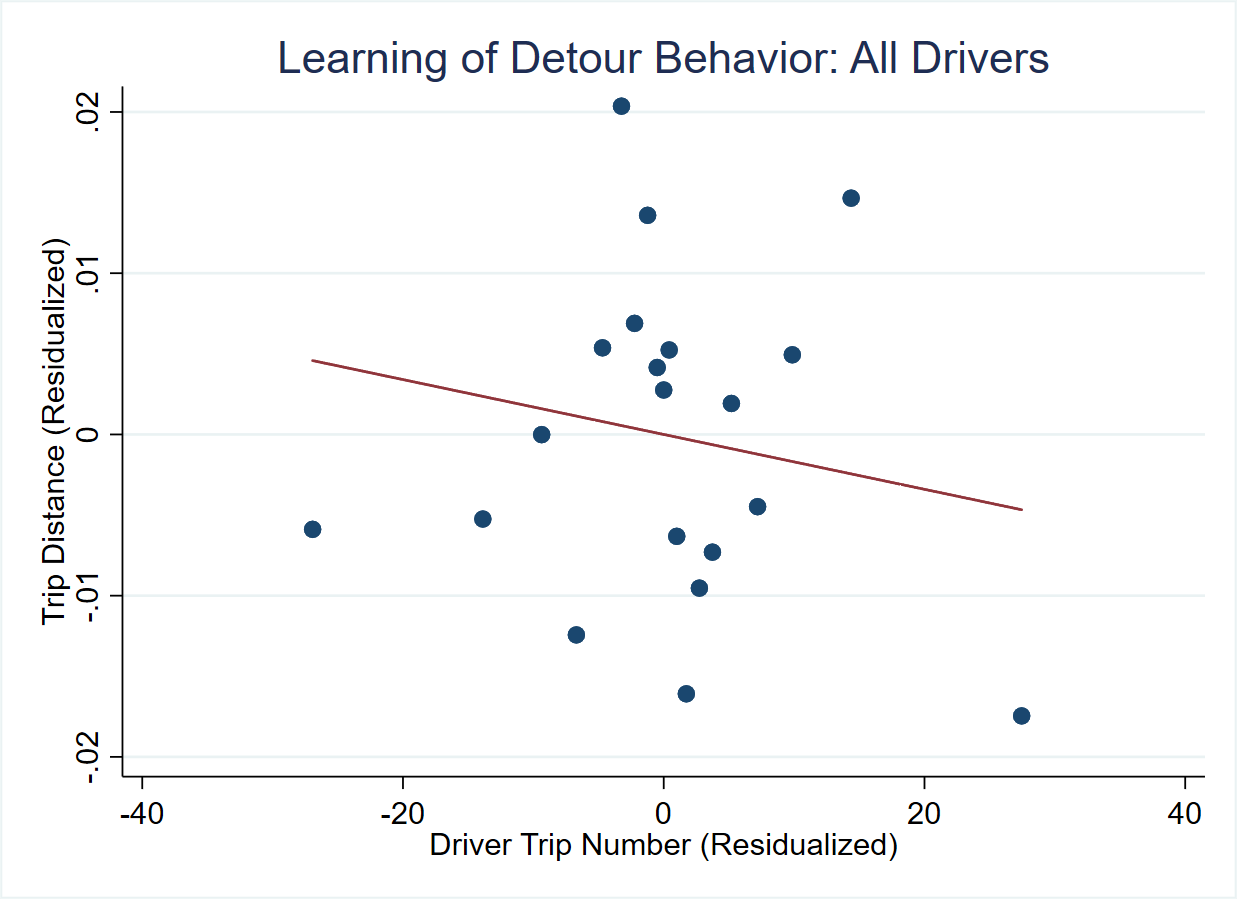}
    \label{learning_plot_fullsample_1}
\end{figure}

\begin{figure}[H]
    \centering
    \caption{Learning of Detour Behaviors: New Drivers}
    \includegraphics[height=280pt]{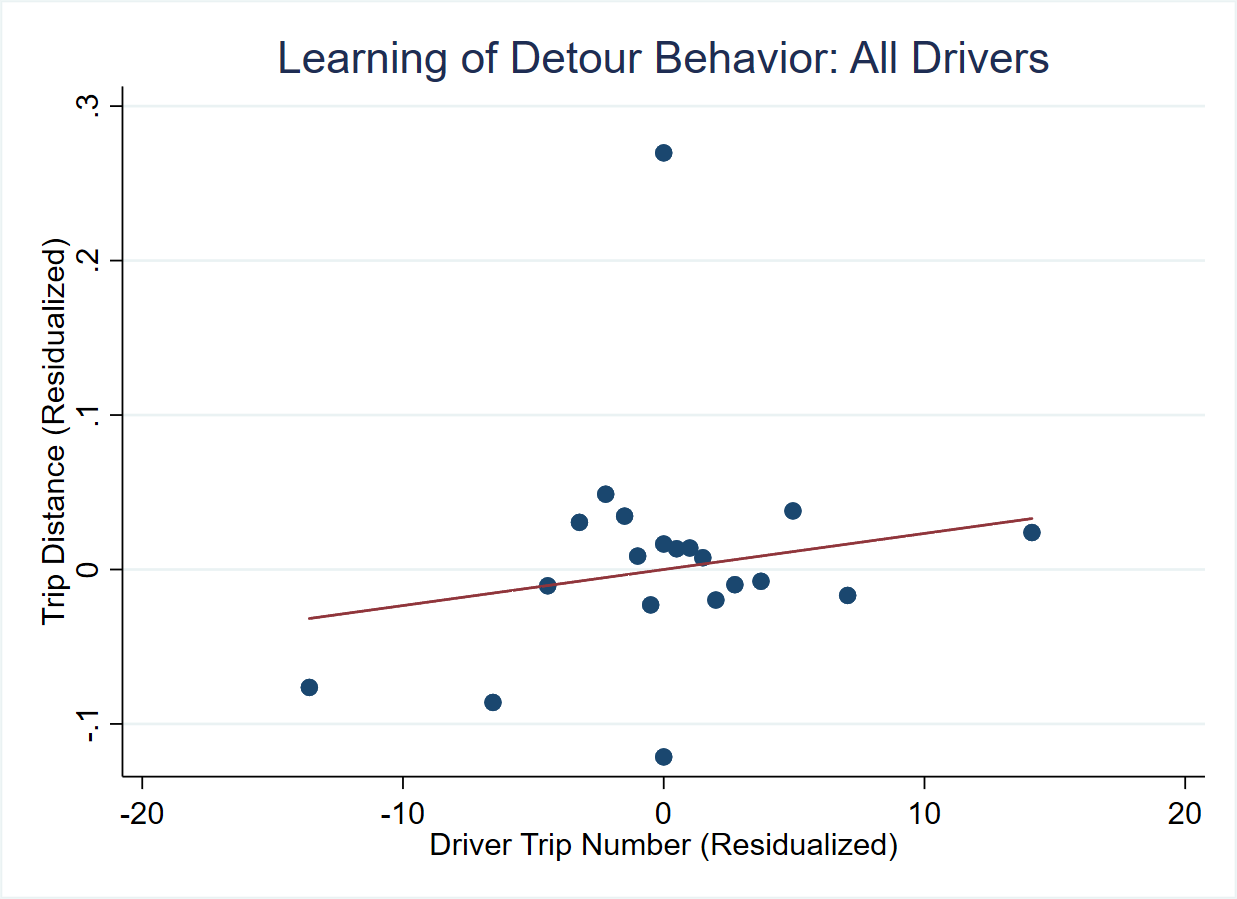}
    \label{learning_plot_newdrivers_1}
\end{figure}

\end{appendices}

\end{document}